\documentclass[12pt]{article}
\pdfoutput=1
\usepackage{amssymb,amsmath,xcolor,empheq}
\numberwithin{equation}{section}
\usepackage{epsfig}
\usepackage{epstopdf}
\usepackage{latexsym}
\usepackage{graphicx}
\usepackage{booktabs}
\usepackage{bbm}
\usepackage{enumitem}
\usepackage[numbers,compress]{natbib}
\usepackage{bm}
\usepackage[T1]{fontenc}
\usepackage{amsthm}

\usepackage{fancybox}

\usepackage[margin=20pt,small]{caption}
\usepackage{subcaption}

\usepackage[toc]{appendix}

\usepackage{color}
\usepackage{datetime}
\usepackage[
      colorlinks=false,
      linkcolor=darkblue,  
      urlcolor=blue,    
      filecolor=blue,     
      citecolor=red,
linktocpage=true,
      pdfstartview=FitV,
      bookmarksopen=true,
	  hidelinks
      ]{hyperref}

\ifpdf
\fi

\newcommand*{\boxedcolor}{black}
\makeatletter
\renewcommand{\boxed}[1]{\textcolor{\boxedcolor}{%
  \fbox{\normalcolor\m@th$\displaystyle#1$}}}
\makeatother

\usepackage{calligra}
\DeclareMathAlphabet{\mathcalligra}{T1}{calligra}{m}{n}

\definecolor{cardinal}{rgb}{0.6,0,0}
\definecolor{darkgreen}{rgb}{0,0.5,0}
\definecolor{golden}{rgb}{0.92, 0.7, 0}
\definecolor{midnight}{rgb}{0, 0, 0.5}
\definecolor{darkblue}{rgb}{0.2, 0, 0.8}

\def\ds{\displaystyle}

\def\cA{{\cal A}}

\def\cF{{\cal F}}

\def\cN{{\cal N}}

\def\cR{{\cal R}}

\def\cT{{\cal T}}

\def\Tr{{\rm Tr}\,}

\def\cL{{\cal L}}

\def\fg{\frak{g}}

\def\cals#1{\mathcal{#1}}

\def\cA{{\cals A}}

\newcommand{\be}{\begin{equation}} \newcommand{\ee}{\end{equation}}
\newcommand{\bea}{\begin{equation} \begin{aligned}} \newcommand{\eea}{\end{aligned} \end{equation}}
\newcommand{\bmu}{\begin{multline}} \newcommand{\emu}{\end{multline}}

\newcommand\equ[1] {\begin{equation}#1\end{equation}}

\newcommand\eqs[1] {\begin{align}#1\end{align}}
\newcommand\eqss[1] {\begin{align}\begin{split}#1\end{split}\end{align}} 
\newcommand\eqsn[1] {\begin{align*}#1\end{align*}}
\renewcommand\( {\left(}
\renewcommand\) {\right)}
\newcommand{\mat}[1]{\begin{pmatrix} #1 \end{pmatrix}}
\newcommand{\dvol}{d\mathrm{vol}}
\newcommand{\vol}{\mathrm{vol}}
\usepackage[cmtip,all]{xy}
\newcommand{\longsquiggly}{\xymatrix{{}\ar@{~>}[r]&{}}}

\begin{document}  

\begin{titlepage}
 
\medskip
\begin{center} 
{\Large \bf  Universal RG Flows Across Dimensions and Holography}

\bigskip
\bigskip
\bigskip
\bigskip
\bigskip

{\bf Nikolay Bobev${}^{(1)}$ and P. Marcos Crichigno${}^{(2)}$   \\ }
\bigskip
\bigskip
${}^{(1)}$
Instituut voor Theoretische Fysica, KU Leuven \\
Celestijnenlaan 200D, B-3001 Leuven, Belgium
\vskip 5mm
${}^{(2)}$ Institute for Theoretical Physics, University of Amsterdam,\\ 
Science Park 904, Postbus 94485, 1090 GL, Amsterdam, The Netherlands
\vskip 5mm
\texttt{nikolay.bobev@kuleuven.be,~p.m.crichigno@uva.nl} \\
\end{center}

\bigskip
\bigskip

\noindent 

\abstract{We study RG flows between superconformal field theories living in different spacetime dimensions which exhibit universal properties, independent of the details of the UV and IR theories. In particular, when the UV and IR theories are both even-dimensional  we establish exact universal relations between their conformal anomaly coefficients.  We also provide strong evidence for similar relations between appropriately defined free energies for RG flows between odd-dimensional theories in the large $N$ limit. Holographically, these RG flows across dimensions are described by asymptotically AdS black branes in a gauged supergravity theory, which we exhibit explicitly. We also discuss the uplift of these solutions to string and M-theory and comment on how the entropy of such black branes is captured by the dual field theory.}

\end{titlepage}


\setcounter{tocdepth}{2}
\tableofcontents

\newpage

\section{Introduction and summary}
\label{sec:intro}

Supersymmetric quantum field theories (QFTs) placed in background fields provide a rich laboratory for testing our understanding of  quantum theories. The  toolbox  for studying such theories has greatly expanded in recent years, leading to a cornucopia of exact results  such as the nonperturbative computation of physical observables, nontrivial tests of holography and other known dualities, and the discovery of many new dualities. In the landscape of consistent supersymmetric QFTs, superconformal field theories (SCFTs) play a distinguished role. Their enhanced symmetry offers greater calculational control and they serve as anchors around which to structure our understanding of the renormalization group (RG) flow. SCFTs in background fields will be the main objects of interest in this work.

Consider placing an SCFT on a curved manifold $M_{d}$. If the manifold is equipped with a conformally flat metric,  superconformal symmetry is preserved (up to well-understood anomalies in even dimensions). On a general curved manifold, however, both supersymmetry and  conformal symmetry are generically broken, which leads to a reduced computational control. This can be remedied by employing a simple and powerful idea, due to Witten \cite{Witten:1988ze}. Loosely speaking, the basic observation is that for supersymmetric QFTs with a continuous global R-symmetry one can turn on a background gauge field for this symmetry, and tune its magnitude so as to cancel (part) of the curvature of the manifold. This procedure, dubbed a topological twist, ensures that there is a particular covariantly constant spinor defined on the curved background, which can be used as a supersymmetry generator.\footnote{There are more general ways to place a supersymmetric QFT on a curved manifold, see for example \cite{Festuccia:2011ws} for a systematic approach. In this work we focus on the topological twist.}

While Witten's idea was originally used  to obtain a topological QFT on $M_{d}$, here we will be interested in a different application, possible when the manifold has a product structure of the form $M_{d}=\Bbb R^{p}\times M_{d-p}$, where $\Bbb R^{p}$ is flat Euclidean space and $M_{d-p}$ is a general curved, smooth, compact and orientable manifold.\footnote{For concreteness, here we focus on QFTs in Euclidean signature, although the discussion is also valid for Lorentzian theories, and we later switch between both signatures. We assume that $M_{d-p}$ is smooth and compact, in particular there are no boundaries, punctures, or other defects, but it should be possible to relax this requirement. } In this case one needs to  perform a topological twist  on $M_{d-p}$ only, a procedure often referred to as a partial topological twist. Then, at length scales much larger than the one set by $M_{d-p}$, the effective dynamics is controlled by a {\it non-topological} supersymmetric theory on $\mathbb{R}^{p}$. This procedure can be interpreted as an RG flow across dimensions, triggered by the local operators in the $d$-dimensional UV SCFT which are turned on by the background fields implementing the  twist.  This procedure is often applied in the literature to study specific QFTs, leading to many new insights into the physics of the resulting $p$-dimensional theory at low energies, as well as to unexpected dualities between this theory and a topological theory on $M_{d-p}$; see \cite{Teschner:2014oja} for a recent review and further references. 

If, in addition to a continuous R-symmetry, the  $d$-dimensional  SCFT in the UV has continuous flavor symmetries, one is free to turn background values for the corresponding flavor gauge fields, without breaking any additional supersymmetry. While this freedom often leads to a large and interesting zoo of $p$-dimensional SCFTs in the IR (see  \cite{Bah:2011vv,Bah:2012dg,Benini:2012cz,Benini:2013cda,Bobev:2014jva,Benini:2015bwz,Kutasov:2013ffl,Kutasov:2014hha,Benini:2015eyy,Hosseini:2016tor,Hosseini:2016ume,Amariti:2017iuz} for a selection of recent references), it is clear that the details of such constructions depend on the particular choice of UV SCFT and background flavor fields. In contrast, any QFT has a stress-energy tensor which, for the SCFTs of interest here, sits in the same supermultiplet with the superconformal R-symmetry current.  The universality of this multiplet structure suggests that twisted compactifications  involving only the metric and the exact superconformal R-symmetry gauge field posses special properties, common to all $d$-dimensional UV SCFTs with a given amount of supersymmetry. Indeed, as we demonstrate below, this expectation bears out. 

A main objective of this work is  to establish  universal properties of this class of RG flows.  Since SCFTs do not exist in dimension greater than six, we take  $d\leq 6$. Since we are interested in obtaining dynamical theories on $\Bbb R^{p}$, we consider $ p\geq 1$, and since no topological twist is required on $S^{1}$ we  take $p \leq  d-2$.  The tools we employ to study this setup are general and non-perturbative. Since the construction is clearly strongly dependent on the identification of the correct superconformal R-symmetry, we make 
use of the various maximization/extremization principles which determine the exact superconformal R-current \cite{Intriligator:2003jj,Jafferis:2010un,Benini:2013cda,Benini:2012cz}. Another powerful tool we use, when both $p$ and $d$ are even, is 't Hooft anomaly matching. As we discuss in detail below, this allows us to establish universal relations between conformal anomalies in the UV and IR theories. In the case of 3d SCFTs placed on Riemann surfaces (i.e., $d=3,\,p=1$) we use recent supersymmetric localization results to identify a universal relation between certain supersymmetric partition functions. Finally, another important weapon in our arsenal is the holographic duality. We now proceed to present a short summary of the main results of our work.

\subsection{Main results}
\label{sec:Basic Idea and Main Results}

Consider a $d$-dimensional SCFT with a continuous R-symmetry and, possibly, global flavor symmetries. The corresponding currents organize themselves into supermultiplets of the form:
\eqs{\label{currentMultiplets}
\{T_{\mu\nu},J_{\mu}^{R},\cdots\}\,, \qquad \{J_{\mu}^{F},\cdots\}\,,
}
where $T_{\mu \nu}$ is the stress-energy tensor, $J_{\mu}^{R}$ is the {\it exact} superconformal R-symmetry current, and $J_{\mu}^{F}$ are flavor currents. The ellipsis stand for other possible bosonic and fermionic operators which depend on the number of supercharges and the dimension of spacetime. While the stress-energy tensor multiplet is omnipresent in  SCFTs, the flavor symmetry currents depend on the specific theory under consideration. As described above, the topological twist is implemented by giving the corresponding sources, $g_{\mu\nu}, A_{\mu}^{R},A_{\mu}^{F}, \cdots$, nontrivial background values. This procedure triggers an RG flow which depends on the details of the UV theory and in general requires a case-by-case analysis. The basic observation we make here is that if  the topological twist is implemented only by turning on background fields which couple to the stress-energy tensor multiplet, i.e. set $A_{\mu}^{F}=0$, the resulting RG flow exhibits universal properties. Based on this observation, we make the following definition: {\it  Given a $d$-dimensional UV SCFT with a certain number of supercharges and continuous R-symmetry, placed on $\Bbb R^{p}\times M_{d-p}$, we define the universal twist of the theory as the partial topological twist on  $M_{d-p}$ along the exact UV superconformal R-symmetry, preserving the maximal possible number of supercharges.} We note that for theories with a large amount of supersymmetry there may be more than one universal twist and we discuss this in detail below. For theories with rational central charges one may find constraints on the topology of the compactification manifold $M_{d-p}$ due to Dirac quantization of the background gauge field (see Section \ref{subsec:toptwist}).\footnote{For theories with an Abelian R-symmetry the R-charges may be irrational which renders the universal twist impossible. These subtleties do not arise if the UV theory has a non-Abelian R-symmetry.}

As is clear by now, the exact superconformal R-symmetry plays a crucial role in our story. Upon a generic (non-universal) twisted compactification of the theory, the exact R-symmetry of the IR SCFT may differ from the one in the UV, due to possible mixing of Abelian R-symmetries with Abelian flavor symmetries along the flow.\footnote{Here we shall assume that the resulting $p$-dimensional theory flows to an interacting SCFT in the IR.} Although the precise mixing may be determined by an appropriate extremization principle, the result depends on the details of the theory and is thus not universal. In the case of flows between even-dimensional SCFTs with a universal twist, however, one can show that there is no such mixing along the flow.\footnote{There is a minor caveat in the case of 4d $\mathcal{N}=1$ SCFTs compactified on a Riemann surface, which we discuss in Section \ref{N=1 Sigma}} This in turn leads to universal relations between the conformal anomaly coefficients, which we denote by the vector $(a,\vec c)$, of the UV $d$-dimensional theory and the IR $p$-dimensional theory of the form: 
\equ{\label{univ gen}
\(\begin{matrix}
a_p\\\vec c_p
\end{matrix}
\)=\, \mathcal U\(\begin{matrix}
a_d\\\vec c_d
\end{matrix}
\)\,.
}
Here $\mathcal U$ is a matrix that depends only on the topology of the compactification manifold,  $M_{d-p}$,  and the data specifying the partial topological twist, but is independent of the details of the SCFTs at both ends of the RG flow. These relations are exact and do not rely on the existence of a Lagrangian description of the UV or IR SCFTs.

We will argue that this universal behavior is not limited  to RG flows between even-dimensional theories and that similar universal relations exist for flows between theories in various dimensions. For RG flows between odd-dimensional theories one cannot rely on 't Hooft anomalies but in view of the $F$-theorem \cite{Jafferis:2011zi} it is natural to search for universal relations between the round-sphere free energy of the UV and IR SCFTs. Although computing  free energies exactly is much harder than computing anomalies, we are nonetheless able to show that the free energies indeed are related by
\equ{\label{eqFintro} 
F_{S^{p}\times M_{d-p}}=u \, F_{S^{d}}\,,
}
to leading order in $N$, in an appropriate large $N$ limit. Here $u$ is again a universal coefficient, depending only on the topology of $M_{d-p}$ and the topological twist performed, but not the details of the SCFTs. The free energy on the l.h.s. of \eqref{eqFintro} can be thought of as an appropriate free energy of the IR theory on $S^{p}$. This universal relation can also be proven by pure field theory methods for the twisted compactifications of a large class of 3d $\cN=2$ SCFTs on a Riemann surface by an analysis of the corresponding matrix models at large $N$  \cite{ABCMZ} (see Section~\ref{sec:Comments on 3d and 5d SCFTs}). For theories in other dimensions we establish the relation in \eqref{eqFintro} using holography.

As mentioned above, in order to establish the universal relations \eqref{univ gen},  \eqref{eqFintro} one needs to assume that the RG flow ends at an interacting SCFT in the IR. Whether this is actually the case is a nontrivial dynamical question,   difficult  to establish by field theory methods. When the $d$-dimensional SCFT admits a weakly coupled holographic dual, however, one can bring holography  to bear on this question. Indeed,  one way to establish the existence of an interacting superconformal fixed point in the IR (at least in the planar limit) is by constructing a supergravity solution that explicitly interpolates between the UV and IR SCFTs. The holographic description of twisted compactifications was first studied in the foundational work of Maldacena and N\'u\~nez (MN) \cite{Maldacena:2000mw}, which built upon results for D-branes wrapping calibrated cycles \cite{Bershadsky:1995qy} (see \cite{Gauntlett:2003di} for a review and further references). We exploit the same approach in our holographic analysis.

In this holographic setting, it is natural to ask what is the supergravity manifestation  of universal  RG flows across dimensions.  Consider the supergravity fields dual to the operators in \eqref{currentMultiplets}. These also organize themselves into multiplets, of the form
\equ{
\{g_{\mu\nu},A_{\mu}^{R},\cdots \}\,,\qquad \{A_{\mu}^{F},\cdots \}\,,
} 
where the gravity multiplet contains the metric $g_{\mu\nu}$ and the graviphoton $A_{\mu}^{R}$. The gauge fields $A_{\mu}^{F}$ belong to vector multiplets.\footnote{These are dynamical supergravity fields and coincide with the background metric and gauge fields in the boundary field theory discussed below \eqref{currentMultiplets} only at the asymptotically AdS boundary.} Since universal twists involve only operators dual to the gravity multiplet, it is natural to expect that the dynamics of this multiplet is sufficient to capture the corresponding universal RG flow. Indeed, one can restrict to this ``minimal'' gauged supergravity theory\footnote{Note that by ``minimal'' here we mean that we consider only the theory containing the gravity multiplet and no extra matter multiplets, not that the theory has the minimal number of supercharges.} in $(d+1)$ dimensions and construct domain wall solutions with a metric of the form
\equ{
ds^{2}_{d+1}=e^{2f(r)}\,ds^{2}_{\Bbb R^{p+1}}+e^{2g(r)}\,ds^{2}_{M_{d-p}}\,.
}
Here $r$ is the ``holographic'' direction, and the metric is (locally) asymptotic to AdS$_{d+1}$ for large $r$ and approaches AdS$_{p+1}\times M_{d-p}$ for small $r$. The solution is supported also by a nontrivial magnetic flux for the graviphoton through 2-cycles in $M_{d-p}$. The entire spacetime can be thought of as a BPS $(p-1)$-brane living  in AdS$_{d+1}$, interpolating between the SCFT$_{d}$ dual in the UV and the  SCFT$_{p}$ dual in the IR. We identify the black brane solutions corresponding to various twisted compactifications in minimal gauged supergravity for $(d+1)=4,5,6,7$ and describe them in detail in Section~\ref{sec:Gauged Supergravity}. In addition, we show that the field theory universal relation in \eqref{univ gen} and \eqref{eqFintro} are correctly reproduced holographically. Many of these supergravity solutions have been found in the literature before, but their field theory interpretation as universal RG flows has not necessarily been appreciated. 

To make contact with top-down constructions in string and M-theory we also emphasize that these low-dimensional gauged supergravity theories arise as consistent truncations from ten and eleven-dimensional supergravity. This means that it is possible to uplift the universal holographic RG flows to string and M-theory. The choice of the internal manifold determines the details of the particular SCFT dual. For instance, uplifting a five-dimensional solution to IIB SUGRA with $S^{5}$ as internal manifold describes a twisted compactification of $\cN=4$ SYM, while taking the internal manifold to be $Y^{p,q}$ corresponds to a twisted compactification of $\cN=1$ quiver gauge theories of the type discussed in \cite{Benini:2015bwz}.

The holographic perspective not only establishes the existence of the IR fixed point, but also suggests universal relations among quantities for flows between even-dimensional and odd-dimensional SCFTs of the form
\equ{\label{univ gen dim}
a_{p}= u_{aF}\,  F_{S^{d}}\,, \qquad \qquad F_{S^{p}\times M_{d-p}}= u_{Fa} \, a_{d}\,.
}  
Here $F$ is a free energy, $a$ is a conformal anomaly coefficient, and $u_{aF},u_{Fa}$ are again universal coefficients, depending only on the compactification manifold $M_{d-p}$ and the topological twist performed. These are nontrivial and powerful predictions that would be interesting to establish directly in field theory, including finite $N$ corrections. Finally, we should stress that the supersymmetric black branes  realizing these universal RG flows across dimensions have a non-vanishing entropy. We compute these entropies in Appendix \ref{app:entropy} and observe interesting relations with field theory quantities such as conformal anomalies and sphere free energies.

The idea of universal RG flows common to a large class of SCFTs has appeared before, both in a holographic \cite{Corrado:2002wx}  as well as in a purely field-theoretic context \cite{Corrado:2004bz,Tachikawa:2009tt}. In these papers, however, the UV and IR theories live in the same number of space time dimensions. Universal supergravity domain walls dual to holographic RG flows, similar in spirit to the ones studied here, were also discussed in \cite{Benini:2013cda,Bobev:2014jva,Benini:2015bwz,Passias:2015gya,Apruzzi:2015wna}.

The rest of the paper is organized as follows. In Section~\ref{sec:Generalities} we review some background material on 't Hooft and conformal anomalies and topological twists. Section~\ref{sec:Field theory} is devoted to a study of twisted compactifications of field theories in various dimensions with a particular focus on universal relations among conformal anomalies. In Section~\ref{sec:Gauged Supergravity} we present the holographic dual description of these universal twisted compactifications and discuss universal relations among various quantities, such as conformal anomalies and free energies from the holographic perspective. We conclude with a short summary and a discussion of various open problems in Section~\ref{sec:Discussion}. In the three appendices we present our conventions on characteristic classes, a short discussion on the relation between some of our results and the entropy of extremal black branes, as well as an observation of a possible two-dimensional analog of the universal RG flow discussed in \cite{Tachikawa:2009tt}.

\section{Generalities}
\label{sec:Generalities}

We begin by reviewing some general background on anomalies in QFTs and basics of topologically twisted supersymmetric QFTs. Readers familiar with this material may skip to Section~\ref{sec:Field theory}.

\subsection{'t Hooft anomalies}

In even-dimensional QFTs, classical symmetries may become anomalous at the quantum level. Quantum anomalies for local symmetries are forbidden in consistent QFTs. However, global (or 't Hooft) anomalies are not only allowed, but are in fact robust physical observables containing exact information about the theory  (see \cite{Harvey:2005it} for a pedagogical review).  't Hooft anomalies for continuous global symmetries are  packaged efficiently in the anomaly polynomial, $I_{d+2}$, of the theory. This is a gauge-invariant $(d+2)$--form which is a polynomial in characteristic classes for the global symmetries of the theory.

We will consider two-dimensional $\cN=(0,2)$, four-dimensional $\cN=1$, and six-dimensional $\cN=(1,0)$ theories, whose R-symmetry groups are $U(1)_{R}, U(1)_{R}$, and $SU(2)_{R}$, respectively.\footnote{This formalism is also applicable for even-dimensional theories with more supersymmetry and we use it extensively in this context in Section~\ref{sec:Field theory}.} In addition, these theories generically also have flavor symmetries. The anomaly polynomials are given by:
\eqs{
\label{I4}
\text{2d}:\qquad I_4 &= \frac{k_{RR}}{2} \, c_{1}(\cF_{R})^{2}-\frac{k}{24} \, p_{1}(\cT_{2}) +I^{\text{ flavor}}_{4}\,,\\
\label{I6}
\text{4d}:\qquad I_{6} &= \frac{k_{RRR}}{6}\, c_{1}(\cF_{R})^{3} -\frac{k_{R}}{24}\, c_{1}(\cF_{R})\,p_{1}(\cT_{4}) +I^{\text{ flavor}}_{6}\,,\\
\label{I8}
\text{6d}:\qquad I_{8}&=\frac{1}{4!}\,\(\alpha \, c_{2}^{2}(\mathcal F_{R})-\beta \, c_{2}(\mathcal F_{R})\,p_{1}(\cT_{6})+\gamma \, p_{1}^{2}(\cT_{6})+\delta \, p_{2}(\cT_{6})\)+I^{\text{ flavor}}_{8}\,.
}
Here $c_{n}(\cF)$ denotes the $n$'th Chern class of the corresponding bundle and  $p_n(\cT_d)$ denotes the $n$'th Pontryagin class of the tangent bundle of the  manifold on which the theory is placed. See Appendix~\ref{App:CC} for our conventions. The various coefficients multiplying these characteristic classes encode the corresponding 't Hooft anomalies for the energy-momentum multiplet in the theory. For the theories considered here this multiplet always contains the superconformal R-symmetry current. If the QFT at hand admits a Lagrangian description these anomalies can be computed by one-loop Feynman diagrams with insertions of the R-current and the energy-momentum tensor.

In the anomaly polynomials above we have not included anomalies for gauge symmetries since in this work we only study consistent QFTs, where such anomalies are absent. Note that we allow for gravitational anomalies since we are discussing QFTs and thus the metric is treated as a non-dynamical background field. We have not given explicit expressions for flavor anomalies as well as mixed flavor-R-symmetry anomalies. These are schematically encoded in the anomaly polynomial $I_{d+2}^{\text{flavor}}$ in \eqref{I4}-\eqref{I8}. The rationale for doing this is that these anomalies depend on the details of the theory and can be ignored  (under certain mild assumptions, to be discussed in Section~\ref{sec:Field theory}) for the purposes of our discussion.  \\

\subsection{Weyl anomaly}
\label{subsec:Weyl}

Another important anomaly for our story is the Weyl (or conformal) anomaly, which captures the failure of the stress-energy tensor to be traceless when an even-dimensional CFT is placed in a nontrivial curved background. Ignoring conventional normalizations the anomaly has the form:
\equ{\label{trace T}
\langle T^{\mu}_{\,\mu}\rangle\sim a \, E_{d}+\sum_{i} c_{i}\,W_{i}\,,
}
where $E_{d}$ is the Euler density in $d$ dimensions and the $W_{i}$ are a set of local, independent, Weyl invariants of the manifold on which the theory is placed. The number of independent invariants of this type depends on the spacetime dimension. There are none in two dimensions, one in four dimensions, and three in six dimensions. 

In a superconformal theory, the stress-energy tensor  and the superconformal R-symmetry current sit in the same supermultiplet. As a consequence, the Weyl and R-symmetry anomaly coefficients in \eqref{I4}-\eqref{I8} and \eqref{trace T} are related by supersymmetry through Ward identities.  In 2d (see for example \cite{Boucher:1986bh}) and 4d \cite{Anselmi:1997am} these relations are:
\begin{equation}
\begin{split}
\label{R-conf anom}
c_r &= 3k_{RR} \;,\qquad c_r - c_l = k \;, \\ 
a_{4d} &= \frac{9}{32} \,k_{RRR} - \frac{3}{32} \,k_R \;,\qquad c_{4d} = \frac{9}{32}\, k_{RRR} - \frac{5}{32}\,k_R \;,
\end{split}
\end{equation}
where $R$ denotes the exact superconformal R-symmetry. 

In 6d there are three tensor structures $W_{i}$ and thus three $c$-type coefficients (see e.g. \cite{Bastianelli:2000hi} and references therein for details). For theories with $\cN=(1,0)$ supersymmetry there is a linear relation\footnote{As argued in \cite{Kulaxizi:2009pz} the relation imposed by supersymmetry is $6c^{(3)}_{6d}+c^{(1)}_{6d}-2c^{(2)}_{6d}=0$.}  among the $c_{i}$'s and thus the independent Weyl anomaly coefficients in 6d can be taken to be $(a_{6d},c^{(1)}_{6d},c^{(2)}_{6d})$.  The expression for the $a$-anomaly coefficient in terms of R-symmetry anomalies was found in \cite{Cordova:2015fha}. The expression for the $c^{(1)}_{6d},c^{(2)}_{6d}$ coefficients was recently determined in \cite{Beccaria:2015ypa,Yankielowicz:2017xkf,Beccaria:2017dmw}. The result is given by the following formulae
\eqs{\label{a6d}
a_{6d}&=\frac27\(8\alpha-8\beta+8\gamma+3 \, \delta\)a_{T}\,,\nonumber\\ 
c^{(1)}_{6d}&=\frac{256}{7}\( 6\alpha-7\beta +8 \gamma+4\delta\)a_{T}\,,\\
c^{(2)}_{6d}&=\frac{64}{7} \( 6\alpha -5 \beta +4 \gamma + 5\delta\)a_{T} \,,\nonumber
}
where $a_{T}$ is the value of $a_{6d}$ for the free $\cN=(2,0)$ tensor multiplet.\footnote{In  \cite{Beccaria:2015ypa} this was normalized to $a_{T}=-\frac{7}{1152}$.  Here we follow the conventions in \cite{Cordova:2015fha} and set $a_{T}=1$.}  In the special case of $\cN=(2,0)$ theories one has the relations   \cite{Beccaria:2015ypa}, $ \gamma=\frac{\beta}{4}$ and $\delta=-\beta$  which imply that $c^{(1)}_{6d}=4c^{(2)}_{6d}$ and there is only one independent $c$-type coefficient, which we take to be $c_{6d}\equiv \frac{7c^{(1)}_{6d}}{384}=\frac{7c^{(2)}_{6d}}{96}$.\footnote{Compared to \cite{Beccaria:2015ypa} we use the normalization $c_{6d}=\frac74 c$ so that the $c$-type coefficient for the free tensor multiplet is 1.} The relations \eqref{a6d} for $\cN=(2,0)$ SCFTs then simplify to 
\equ{\label{ac6dalphabeta2,0}
a_{6d}=\frac{16}{7}\(\alpha-\frac98\,\beta\)\,,\qquad c_{6d}=4\(\alpha-\frac32\, \beta\)\,.
}
For $\cN=(2,0)$ SCFTs in the ADE class one has
\equ{
\alpha=d_{G}h_{G}+r_{G}\,, \qquad \beta=4\gamma=-\delta=\frac{r_{G}}{2}\,,
}
where $d_{G}$, $r_{G}$, and $h_{G}$ are the dimension, rank, and Coxeter number of the  group $G$, respectively, satisfying the group theory identity $d_{G}=r_{G}(1+h_{G})$. We note that for these theories $\frac47\leq a_{6d}/c_{6d}\leq1$, the lower bound being saturated in the large $N$ limit for the $A_N$ and $D_N$ theories. 

\subsection{Topological twist}
\label{subsec:toptwist}

If a supersymmetric QFT in flat space is placed in a general background for the metric and gauge fields, supersymmetry will  be broken. It was recently understood how to systematically arrange the background fields in such a way that some amount of supersymmetry is preserved \cite{Festuccia:2011ws}. Here we will specialize to one particular such way, introduced by Witten in \cite{Witten:1988ze} and known as a topological twist. The basic idea can be summarized as follows. If  the QFT at hand has a continuous R-symmetry, one can turn a background gauge field, $A_{\mu}^{R}$, that couples to the R-symmetry current. One can then adjust the magnitude of the  background field so as to cancel the nontrivial part of the spin connection $\omega_{\mu}$ on the curved manifold. Schematically, one tunes $A_{\mu}^R = -\frac{1}{4}\,\omega_{\mu}$ so that the generalized Killing spinor equation takes the  form 
\begin{equation}\label{genkillingspinor}
\tilde{\nabla}_{\mu}\, \epsilon = \(\partial_{\mu}+\frac14 \, \omega_{\mu}+A_{\mu}^{R}\)\,\epsilon=\partial_{\mu}\epsilon=0\,.
\end{equation}
The last equation in \eqref{genkillingspinor} admits a constant spinor solution on any spin manifold, implying that some amount of supersymmetry is preserved in this nontrivial background for the metric and R-symmetry gauge field.\footnote{Here we have been schematic, omitting Lorentz and R-symmetry indices. If these are included one sees that the cancellation of the spin connection by the background R-symmetry can occur, at most, when acting on half of the components of $\epsilon$ and thus, at most, half supersymmetry can be preserved in this way.} It is clear that this method for preserving some supersymmetry can work only if the nontrivial part of the spin connection can be embedded in the R-symmetry group of the QFT in flat space.\footnote{We are being slightly imprecise here. In general, one needs to cancel only part of the spin connection on the curved manifold if the spin connection after the twist admits covariantly constant spinors. For example, on K\"ahler manifolds in four real dimensions one may cancel only the $U(1)$ part of the $U(2)=U(1)\times SU(2)$ structure group.}

The case of interest to us here is when $M_{d}$ is a product manifold of the form $M_{d}=\Bbb R^{p}\times M_{d-p}$, with $M_{d-p}$ a curved, compact, smooth manifold. Then, the topological twist is performed only along  $M_{d-p}$ and thus at low energies (compared to the scale set by the size of $M_{d-p}$) one expects to have a physical, i.e., non-topological, supersymmetric theory on $\mathbb{R}^{p}$. This is known as a partial topological twist.\footnote{In the original construction of Witten the manifold $M_d$ did not necessarily have a product structure with a flat factor and thus the resulting theory on the curved manifold was topological. Although also interesting, we do not consider such theories here.  
} An important point for many of our constructions below is that this topological twist is naturally realized in sting theory on the world-volume of D- and M-branes wrapping calibrated cycles in special holonomy manifolds \cite{Bershadsky:1995qy}. 

If, in addition to a continuous R-symmetry, the QFT at hand has a continuous flavor symmetry,  one can turn a more general background $ A_{\mu}^{R\, '} =A_{\mu}^{R}+\sum_{i}a_{i}\,A_{\mu}^{F_{i}}$, where  $A_{\mu}^{F_{i}}$ are background fields for the flavor symmetry and the $a_{i}$  are free parameters. Since, by definition, the supersymmetry parameter $\epsilon$ is not charged under flavor symmetries, the Killing spinor Equation \eqref{genkillingspinor} is not modified by turning on background flavor fields  and the amount of supersymmetry preserved is  unchanged. When the R-symmetry group is Abelian this freedom reflects the fact that the R-symmetry is ambiguous, as any linear combination of a ``reference'' R-symmetry with Abelian flavor symmetries is again an R-symmetry. Although one is free to choose any  reference R-symmetry one likes,  in the case of SCFTs there is a preferred R-symmetry, namely the {\it superconformal R-symmetry}, $R_{SC}$, whose corresponding current belongs to the stress-energy tensor supermultiplet. This unique R-symmetry can be determined  by maximization/extremization principles in any integer dimension in the range $1\leq d\leq4$ \cite{Intriligator:2003jj, Jafferis:2010un,Benini:2013cda,Benini:2012cz,Benini:2015eyy}.  For SCFTs it is thus natural to take the superconformal R-symmetry to be the reference R-symmetry and write the background field as:
\equ{\label{general R flavor}
A_{\mu}^{R }=A_{\mu}^{R_{\text{SC}}}+\sum_i a_i\, A_{\mu}^{F_{i}}\,.
}
It is then easy to understand why the universal RG flows across dimensions studied in this paper are special; these correspond to setting $a_{i}=0$, i.e.,  the special choice of background gauge field which extends only along the exact superconformal R-symmetry in the UV. Although not the focus here, flows for generic values of $a_i$ are of course also interesting, and lead to a plethora of RG flows across dimensions; see for example \cite{Bah:2011vv,Bah:2012dg,Benini:2012cz,Benini:2013cda,Bobev:2014jva,Benini:2015bwz,Benini:2015eyy,Benini:2016rke}.

There is an important subtlety to keep in mind when performing topological twists. For any gauge-invariant operator $\mathcal O$ in the theory one must impose the Dirac quantization condition
\equ{\label{QC}
\frac{1}{2\pi}\, \Tr \int_{C_2}  F_{R}\cdot \mathcal O=n\, \mathcal O\,, \qquad n\in \Bbb Z\,,
}
where $ F_{R}$ is the background R-symmetry curvature for the background field \eqref{general R flavor}  and $C_2$ is any compact 2-cycle in $M_{d}$. As a consequence, for the universal topological twist to be well defined, the exact superconformal R-charge of all gauge-invariant operators in the theory must be a rational number. This may not be the case in some theories with four Poincar\'e supercharges, such as 3d $\mathcal{N}=2$ and 4d $\mathcal{N}=1$ SCFTs, in which case the universal topological twist is ill-defined. Nonetheless, one can easily find an infinite number of such SCFTs with rational R-charges, so this is not an important obstruction to discuss universal properties of these constructions. For theories with rational R-charges the quantization condition in \eqref{QC} may lead to constraints on the topology of $M_{d-p}$ and we shall discuss such cases below (see Table~\ref{tab:table1} and the discussion above it). As emphasized in \cite{Benini:2015bwz}, in such situations one can circumvent these constraints on the universal topological twist by including flavor magnetic fluxes $a_i$, and adjusting them in a way consistent with \eqref{QC}. This procedure, however, is theory-specific and thus not universal. In this paper we restrict ourselves to SCFTs, and choices of manifolds $M_{d-p}$, for which the quantization condition \eqref{QC} is satisfied when all flavor fluxes are set to zero. This excludes, in particular, theories with irrational R-charges.

\section{Field theory}
\label{sec:Field theory}

As outlined in the Introduction, the main characters in our story are superconformal field theories with a continuous R-symmetry, which we place on a manifold of the form
$M_{d}=\Bbb R^{p}\times M_{d-p}$ with a partial topological twist on $M_{d-p}$. In this section, we consider SCFTs with different number of supercharges and various values of $d$ and $p$. Our goal  is to extract some physical information of the low-energy  effective theory on $\mathbb{R}^p$ at the end of the RG flow.

For flows between even-dimensional SCFTs (both $d$ and $p$ even) the basic tools that allow us to establish the universal relations \eqref{univ gen}  are anomaly matching  and superconformal symmetry. The calculation proceeds along the lines of the analysis in \cite{Alday:2009qq,Benini:2009mz,Bah:2011vv,Bah:2012dg,Benini:2012cz,Benini:2013cda,Bobev:2014jva,Benini:2015bwz}. Since this will be used repeatedly throughout this section, let us summarize the general strategy  before  studying different cases. We start with a SCFT$_{d}$ with an anomaly polynomial $I_{d+2}$. Performing a partial topological twist on $M_{d-p}$ modifies the global symmetry bundles, leading to a new anomaly polynomial $I_{d+2}^{\text{twisted}}$, which we then integrate over $M_{d-p}$ to arrive at the $I_{p+2}$ anomaly polynomial of the $p$-dimensional IR theory
\equ{\label{intpoly}
I_{p+2}=\int_{M_{d-p}}I_{d+2}^{\text{twisted}}\,.
}
This equation determines the R-symmetry anomalies in the IR in terms of those in the UV, encoding the 't Hooft anomaly matching condition  \cite{'tHooft:1979bh}. Assuming the UV and IR theories are superconformal we can then use superconformal Ward identities at the two fixed points to express the R-symmetry anomalies in terms of conformal anomalies. This ultimately leads to the universal relations of the form \eqref{univ gen} among IR and UV central charges.

For situations in which either $d$ or $p$ is odd, one cannot rely on anomaly matching, making it harder to analyze the resulting RG flow. As argued in the Introduction, a natural  quantity to consider in the absence of anomalies is an appropriate supersymmetric partition function, or free energy, of the CFT.\footnote{See \cite{Giombi:2014xxa} for a somewhat related proposal on how to interpolate between CFTs in even and odd dimensions.} This is a much harder task, not only because  the computation of such partition functions is technically more involved, but also because it is not  obvious how to approach such calculations in a universal way, i.e., without referring to a specific theory.  Some progress on this hard question has been made recently for twisted compactifications of 3d $\cN=2$ theories on a Riemann surface $\Sigma_{\fg}$  ($d=3$, $p=1$) in the planar limit.  In this case one indeed finds a universal relation between the supersymmetric three-sphere free energy, $F_{S^{3}}$, of the UV 3d SCFT and a certain topologically twisted partition function, $F_{\Sigma_{\fg}\times S^{1}}$, which is identified with a Witten index of the effective 1d theory in the IR \cite{ABCMZ} (see also \cite{BMP}). This relation has been established explicitly for a large class of quiver gauge theories, to leading order in $N$, and we discuss it in more detail in Section~\ref{sec:Comments on 3d and 5d SCFTs}.

\subsection{6d SCFTs}
\label{sec:6d SCFTs}

We begin our exploration of RG flows across dimensions from $d=6$, the maximal dimension in which a SCFT can exist \cite{Nahm:1977tg}.  We will consider theories with both $\cN=(1,0)$ and $\cN=(2,0)$ supersymmetry. 

\subsubsection{$\cN=(1,0)$ }
\label{sec:N=(1,0) theories}

SCFTs with $\cN=(1,0)$ supersymmetry have eight real supercharges\footnote{Here we count only real components of Poincar\'e supercharges. When we have a conformal theory there is as usual the accompanying superconformal supercharges.} and an $SU(2)_{R}$ R-symmetry. They may also have global flavor symmetries. The study of these theories has recently attracted much attention and a general formula for their anomaly polynomial was derived in \cite{Ohmori:2014kda,Intriligator:2014eaa}. The relation between 't Hooft and Weyl anomalies imposed by superconformal Ward identities was found in \cite{Cordova:2015fha,Beccaria:2015ypa,Yankielowicz:2017xkf,Beccaria:2017dmw}.  We now study these theories with a partial topological twist on Riemann surfaces and on K\"ahler four-manifolds. 

\paragraph{On Riemann surfaces.}

Consider a smooth Riemann surface $\Sigma_{\fg}$ with holonomy group $U(1)_{\Sigma}$. There is a unique way to embed $U(1)_{\Sigma}$ into $SU(2)_{R}$ while preserving minimal 4d $\cN=1$ supersymmetry. At the level of line bundles, the topological twist amounts to the replacement:
\equ{
\mathcal F_{R}^{(6d)}\to \mathcal F_{R}^{(4d)}-\frac{\kappa}{2} \,t_{\fg}\,,
}
where $t_{\fg}$ is the Chern class of the tangent bundle to $\Sigma_{\fg}$, normalized as in \eqref{norm tg}, and the coefficient $-\kappa/2$ is fixed by supersymmetry, where $\kappa$ is the normalized curvature of $\Sigma_\fg$ defined in \eqref{hdef}.  Implementing the twist in the anomaly polynomial \eqref{I8}, integrating over $\Sigma_{\fg}$, and using \eqref{norm tg} one finds 
\equ{\label{I8p} 
\int_{\Sigma_{\fg}}  I_{8}^{\text{twisted}}=\frac{1}{12}\, (\fg-1)\,\(\,2\alpha \,c_{1}(\mathcal F_{R})^{3}-\beta \, c_{1}(\mathcal F_{R})\,p_{1}(T)\,\)+I_{6}^{\text{flavor}}\,.
}
Comparing this  to \eqref{I6} we read off the resulting 4d 't Hooft anomaly coefficients
\equ{\label{rrr r 6d 4d}
k_{RRR}=(\fg-1)\, \alpha\,, \qquad k_{R}=2(\fg-1)\,\beta\,.
}
Many $\cN=(1,0)$ SCFTs have non-Abelian flavor symmetry groups and thus we will assume that the $U(1)_R$ superconformal R-symmetry of the IR 4d theory is the same as the Cartan subgroup of the UV $SU(2)_R$ preserved by the topological twist.\footnote{It would be nice to put this statement on a firmer footing by analyzing the general 6d $\cN=(1,0)$ anomaly polynomials of \cite{Ohmori:2014kda,Intriligator:2014eaa}.} With this identification, we can use \eqref{R-conf anom}  to find the following expression for the 4d Weyl anomalies
\eqs{\label{2x2matrix a4dc4d}
\text{4d $\cN=1$:}&&\begin{pmatrix}
a_{4d}\\
c_{4d}
\end{pmatrix}= \frac{(\fg-1)}{32}\begin{pmatrix}
9& -6\\
9&-10
\end{pmatrix} \begin{pmatrix}
\alpha\\
\beta
\end{pmatrix}\,.
}
Note that the 4d 't Hooft and Weyl anomalies depend only on $\alpha,\beta$ and not on the purely gravitational anomalies $\gamma,\delta$. Recall that the 6d Weyl anomaly coefficients are given in terms of 't Hooft anomalies in \eqref{a6d}. Since there are three independent Weyl anomaly coefficients in $\cN=(1,0)$ theories, but four R-symmetry anomalies, one invert the relations \eqref{a6d} to write \eqref{2x2matrix a4dc4d} as a relation purely among 6d and 4d Weyl anomalies. It is possible, however, to do so in the case of a $\cN=(2,0)$ theory, which we discuss below. We note that requiring that the ratio $a_{4d}/c_{4d}$  satisfies the Hofman-Maldacena bound $\frac12\leq a_{4d}/c_{4d}\leq \frac32$ imposes conditions on the values of $\alpha,\beta$. We discuss Hofman-Maldacena bounds in more detail in Section~\ref{sec:Comments on Hofman-Maldacena bounds}. 

If the SCFT at hand admits a suitable large $N$ limit the pure R-symmetry anomaly dominates over gravitational anomalies, i.e., $\alpha>>(\beta, \gamma, \delta)$ in \eqref{a6d} in which case the relation \eqref{2x2matrix a4dc4d} becomes
\equ{\label{univ rel 64 large N}
c_{4d}\simeq a_{4d}\simeq (\fg-1)\,\frac{63}{512} \, a_{6d}\,.
}
Assuming $a_{6d}>0$, a positive central charge in 4d is obtained only for $\fg>1$. We will derive this universal ratio from holography in  Section~\ref{sec:Minimal 7d gauged supergravity}. 

Finally we note that the universal relation in \eqref{2x2matrix a4dc4d} is satisfied for the particular models of twisted compactifications of six-dimensional $(1,0)$ SCFTs discussed in Section 7 of \cite{Razamat:2016dpl}, see in particular Equations (7.1) and (7.9).

\paragraph{On K\"ahler four-manifolds.}
\label{sec:6d on Kahler}

Consider a K\"ahler four-manifold $M_{4}$, whose holonomy group is (contained into) $U(2)_{s} = SU(2)_{s}\times U(1)_{s}$. We denote the first Pontryagin number and the Euler number of $M_{4}$ by $P_{1}$ and $\chi$, respectively. These can be written in terms of the  Chern roots $t_{1,2}$  of the tangent bundle to $M_{4}$ as 
\equ{
P_{1}=\int_{M_{4}}(t_{1}^{2}+t_{2}^{2})\,, \qquad \chi=\int_{M_{4}} t_{1}\,t_{2}\,.
}
To preserve 2d $\cN=(0,2)$ supersymmetry we turn on a background for the Cartan of the 6d $SU(2)$ R-symmetry, proportional to the $U(1)_{s}$ spin connection.\footnote{One may also consider turning on a background gauge field proportional to the $SU(2)_{s}$ spin connection. For 6d theories with only $(1,0)$ supersymmetry this results in a 2d theory with $(0,1)$ supersymmetry and thus no continuous R-symmetry. Our anomaly matching procedure is thus not applicable and we do not consider this case further here. For $\cN=(2,0)$ SCFTs this twist leads to a 2d theory with $(0,2)$ supersymmetry and was studied in Section 5.1 of \cite{Gauntlett:2000ng} as well as Section 6.3  of \cite{Benini:2013cda}.} This amounts to 
\equ{
\mathcal F_{R}^{(6d)}\to \mathcal F_{R}^{(2d)}+\frac12 \, t_{1}+\frac12 \, t_{2}\,.
}
Making this replacement in the anomaly polynomial \eqref{I8}, using the relations in \eqref{p1p2M2M4}, and integrating the twisted anomaly polynomial over $M_{4}$ leads to the 2d 't Hooft anomalies 
\equ{
k_{RR}=\frac18\, \alpha \(P_{1}+2\chi\)-\frac{1}{12}\,\beta\,P_{1}, \qquad k=\frac14\,\beta \(P_{1}+2\chi\)-(2\gamma +\delta)\, P_{1}\,.
}
Thus, the central charges at the 2d fixed point are given by 
\eqss{\label{2x2matrix clcr M4}
c_{r}&=\frac{3}{8} \,\alpha\,(P_{1}+2\chi)-\frac{1}{4}\,\beta\,P_{1}\,,\\
c_{l}&=\frac14\,\(\frac{3}{2} \, \alpha-\beta\)(P_{1}+2\chi)-\frac14\,\(\beta -8\gamma-4\delta\)\,P_{1}\,.
}
The same result for the two-dimensional conformal anomalies was derived recently in \cite{Apruzzi:2016nfr} (see in particular Equations (2.36)-(2.37) in \cite{Apruzzi:2016nfr}). As we shall see in Section~\ref{sec:Minimal 7d gauged supergravity}, the holographic dual of this flow across dimensions exists only when the K\"ahler manifold is negatively curved. In this case we can use the following relation between the topological invariants and the volume of the manifold identity\footnote{See Appendix H of \cite{Benini:2013cda} for a short summary on some relevant facts on four-manifolds.}
\equ{\label{volM4neg}
P_{1}+2\chi=\frac{1}{2\pi^{2}}\,\vol(M_{4})\,. 
} 
Using this, together with the fact noted above that $\alpha>>(\beta, \gamma, \delta)$ in the holographic limit,  \eqref{2x2matrix clcr M4} becomes
\equ{\label{crcl64largeN}
c_{r}\simeq c_{l}\simeq \frac{21}{256\pi^{2}}\,\vol(M_{4})\,a_{6d}\,,
}
to leading order in $N$. We will derive this universal ratio holographically in Section~\ref{sec:Minimal 7d gauged supergravity}.

The twisted compactification of a 6d $\cN=(1,0)$ theory on a three-manifold is possible by mixing the $SO(3)\simeq SU(2)$ holonomy group with the $SU(2)_{R}$ R-symmetry. However, in this case one does not expect to obtain a theory with a continuous R-symmetry in the IR and we do not study this case here. For $\cN=(2,0)$ SCFTs, however, one is equipped with an $SO(5)$ R-symmetry group which allows for more general topological twists. We study this next.

\subsubsection{$\cN=(2,0)$ }
\label{subsec:6dN20CFT}

General twisted compactifications of $\cN=(2,0)$ SCFTs on Riemann surfaces and on four-manifolds were studied in \cite{Bah:2011vv,Bah:2012dg} and  \cite{Benini:2013cda}, respectively. Particular twists on Riemann surfaces, which we identify here as universal twists,  were studied in  \cite{Maldacena:2000mw,Benini:2009mz}. Here we  reproduce the results in these references, emphasizing the universal aspects.  

We will focus on two types of topological twists involving only an Abelian background gauge field. The first is the universal twist of the $\cN=(1,0)$ theory described in Section~\ref{sec:N=(1,0) theories} but now applied to $\cN=(2,0)$ theories.  This corresponds to twisting along the diagonal combination of the Cartan $SO(2)_{A}\times SO(2)_{B}\subset SO(5)_R$. The second twist is possible only for $\cN=(2,0)$ theories and preserves twice the amount of supersymmetry. It can be viewed as first decomposing the R-symmetry group in a block-diagonal form as $SO(2)_{A}\times SO(3)_{B}\subset SO(5)_R$ and then twisting along $SO(2)_{A}$.  

\paragraph{On Riemann surfaces.}

We  first discuss the universal twist of the $\cN=(1,0)$ theory, applied to the $\cN=(2,0)$ case.  This twist of the maximal theory on  $\Sigma_{\fg}$ was considered by Maldacena and N\'u\~nez in \cite{Maldacena:2000mw} and we shall refer to it as the MN $\cN=1$ twist. As argued above, \eqref{2x2matrix a4dc4d} still holds in the $\cN=(2,0)$ case, which using \eqref{ac6dalphabeta2,0} can be written as
\eqs{\label{ac4dn=120}
\text{4d $\cN=1$:}&& \begin{pmatrix}
a_{4d}\\
c_{4d}
\end{pmatrix}\,=\frac{(\fg-1)}{384}\begin{pmatrix}
105& -33\\
49&-1
\end{pmatrix} \begin{pmatrix}
a_{6d}\\
c_{6d}
\end{pmatrix}\,.
}
This reproduces the central charges for the 4d $\mathcal{N}=1$ twisted compactifications of the ADE $\cN=(2,0)$ theories derived in  \cite{Benini:2009mz,Bah:2012dg}.

The twist along $SO(2)_{A}$ on $\Sigma_{\fg}$ preserves 4d $\cN=2$ supersymmetry \cite{Maldacena:2000mw} (see also \cite{Witten:1997sc}) and we shall  refer to it as the MN $\cN=2$ twist. It amounts to the following line bundle shifts
\equ{
\mathcal F_{R_{A}}^{(6d)}\to \mathcal F_{R_{0}}^{(4d)}-\frac{\kappa}{2} \,t_{\sigma}\,,\qquad \mathcal F_{R_{B}}^{(6d)}\to \mathcal F_{SU(2)}^{(4d)}\,.
}
Performing the twist in the anomaly polynomial and integrating over $\Sigma_{\fg}$ one finds:
\eqs{\label{matMN2}
\text{4d $\cN=2$:}&& \begin{pmatrix}
a_{4d}\\
c_{4d}
\end{pmatrix}\,=\frac{(\fg-1)}{72}\begin{pmatrix}
21& -6\\
14&-2
\end{pmatrix} \begin{pmatrix}
a_{6d}\\
c_{6d}
\end{pmatrix}\,.
}
Again, this result is compatible with the results in \cite{Gaiotto:2009gz,Bah:2012dg}. In the large $N$ limit the relation between 4d and 6d conformal anomalies becomes
\equ{\label{univ rel 64 large N z=1}
c_{4d}\simeq a_{4d}\simeq\frac{7}{48}\,(\fg-1) \, a_{6d}\,.
}
We will derive this relation holographically in Section~\ref{sec:maximal 7d supergravity}.

\paragraph{On K\"ahler four-manifolds.} 

Applying the universal twist of the $\cN=(1,0)$ theory on $M_{4}$ to  $\cN=(2,0)$ theories, the 2d central charges are still given by \eqref{2x2matrix clcr M4} which using   \eqref{ac6dalphabeta2,0} can be written as
%
\eqss{ \label{c2dn=2020}
c_{r}&=\frac{3}{32}\,(7a_{6d}-3c_{6d})\, (P_{1}+2\chi)-\frac{1}{24}\,(7a_{6d}-4c_{6d})\,P_{1}\,,\\
c_{l}&=\frac{1}{96}\,(35a_{6d}-11c_{6d})\, (P_{1}+2\chi)-\frac{1}{8}\,(7a_{6d}-4c_{6d})\,P_{1}\,.
}
This matches the result obtained in \cite{Benini:2013cda} for ADE theories (see Equation (6.6) there and set $z=\epsilon=0$, which corresponds to the universal twist discussed here).

The second twist, along $SO(2)_{A}$, amounts to
\equ{
\mathcal F_{R_{A}}^{(6d)}\to \mathcal F_{R_{0}}^{(2d)}+\frac{1}{2} \,t_{1}+\frac{1}{2}\, t_{2}\,,\qquad \mathcal F_{R_{B}}^{(6d)}\to \mathcal F_{R_{3}}^{(2d)}\,,
}
and leads to a 2d $(0,4)$ theory with the following anomalies derived in \cite{Benini:2013cda}:
\eqs{ \nonumber
\text{2d $\cN=(0,4)$:}\qquad c_{r}&=\frac{7}{12}\,(P_{1}+2\chi)(c_{6d}-a_{6d})+ \frac{1}{6}\,(P_{1}+3\chi)(7a_{6d}-4c_{6d})\,, \\ \label{c2dn=2004}
 \qquad k&=c_{l}-c_{r}=(P_{1}+\chi)(7a_{6d}-4c_{6d})\,.
}
As discussed in \cite{Benini:2013cda} this 2d SCFTs does not seem to admit a holographically dual AdS$_3$ description at large $N$. One reason for this might be that the theory obtained in this way does not have a normalizable vacuum, like the $\cN=(4,4)$ $\sigma$-model onto the Hitchin moduli space discussed in \cite{Bershadsky:1995vm}. 

Another possible twist on K\"ahler four-manifolds is to turn on a nonabelian R-symmetry background. This corresponds to identifying the $U(2)_s$ spin connection with the $U(2)\subset SO(4)\subset SO(5)_R$ inside the R-symmetry group. This twist preserves 2d $\cN=(1,2)$ supersymmetry and was studied in Section~6.2 of \cite{Benini:2013cda} and we do not discuss it further here. 

\paragraph{On product four-manifolds $M_{4}=\Sigma_{1}\times \Sigma_{2}\,$.} 

When $M_{4}$ is taken to be a product of Riemann surfaces $\Sigma_{1}\times \Sigma_{2}$, with spin connections $\omega_{1,2}$, the holonomy group is reduced to $U(1)_{\Sigma_{1}}\times U(1)_{\Sigma_{2}}$ and there is an additional universal twist possible. We take both Riemann surfaces to have negative curvature for simplicity.  This was studied in \cite{Benini:2013cda}. The twist can be defined by considering the Cartan $SO(2)_{A}\times SO(2)_{B}$ subgroup of the $SO(5)$ R-symmetry and identifying the spin connection $\omega_{1}$ with $SO(2)_{A}$ and the spin connection  $\omega_{2}$ with $SO(2)_{B}$ (or vice-versa). This is the twist studied in Section~3 of \cite{Gauntlett:2001jj}. This preserves 2d $\cN=(2,2)$ supersymmetry and leads to a 2d theory with central charges (see Equation (5.23) in \cite{Benini:2013cda})
\eqs{\label{clcr2022}
\text{2d $\cN=(2,2)$:}&&c_{r}=c_{l}=\frac13 \,(\fg_{1}-1)(\fg_{2}-1)\,(14a_{6d}-5c_{6d})\,. 
}
Using the explicit conformal anomalies for the ADE series of $\cN=(2,0)$ theories one can show that the 2d central charges are an integer multiplet of $3$ which suggests an interpretation of the 2d theory as a nonlinear $\sigma$ model on a Calabi-Yau target space. In the large $N$ limit the central charges become
\equ{\label{c2dn=2022}
c_{r}= c_{l}\simeq \frac74\,(\fg_{1}-1)(\fg_{2}-1)\,a_{6d}\,.
}
We will reproduce this holographically in Section~\ref{sec:maximal 7d supergravity}.

Other twisted compactifications on four-manifolds leading to 2d $\cN=(0,2)$ and $\cN=(0,1)$ theories, and their supergravity duals,  were studied in Section 5 of \cite{Gauntlett:2000ng} (see also \cite{Benini:2013cda}).

\paragraph{On five- and three-manifolds.}

The twisted compactifications of 6d $\cN=(2,0)$ theories on the worldvolume of M5-branes on smooth five- and three-manifolds were discussed in \cite{Acharya:2000mu,Gauntlett:2000ng}. In the former case, the whole $SO(5)_R$ R-symmetry group is turned on to implement the twist, which leads to a superconformal quantum mechanics with a single supercharge (see Section~3.3 in \cite{Gauntlett:2000ng}). In the latter case the twist is obtained by considering $ SO(2)\times SO(3)\subset SO(5)_R$ and twisting along $SO(3)$, which leads to a 3d $\cN=2$ theory (see Section~3.1 in \cite{Gauntlett:2000ng}). These 3d theories were studied also later in \cite{Dimofte:2011ju}.
 
Since we do not have anomalies  at our disposal in 3d and 1d, we are limited in our ability to extract universal information about the RG flow across dimensions with field theory techniques. However, the holographic dual description of these RG flows has been constructed in \cite{Gauntlett:2000ng}, showing that the IR fixed points exist, at least at large $N$. The AdS/CFT dictionary then leads to a universal prediction which would be interesting to test directly in field theory by computing the (partially) twisted partition functions $Z_{S^{1}\times M_{5}}$ and $Z_{S^{3}\times M_{3}}$ and comparing it to the Weyl anomaly coefficients $a_{6d},c_{6d}$.  We discuss this further in Section~\ref{sec:maximal 7d supergravity} below.

\subsection{4d SCFTs}

Here we consider 4d $\cN=1$ and $\cN=2$ SCFTs on a smooth Riemann surface $\Sigma_{\fg}$. In the case of $\cN=1$ theories there is only one possible topological twist, preserving $\cN=(0,2)$ supersymmetry. In the case of $\cN=2$ one may also consider twists preserving $\cN=(2,2)$ and $\cN=(0,4)$. We also comment briefly on twisted compactifications of theories with $\mathcal{N}>2$ as well as compactifications on 3-manifolds.

\subsubsection{$\cN=1$}
\label{N=1 Sigma}

The partial topological twist of $\cN=1$ SCFTs on a compact Riemann surface was discussed in some detail in \cite{Benini:2015bwz}, where an example of the universal relations discussed in this work was presented.  At the level of the R-symmetry bundle the topological twist amounts to the shift:
\be
\label{relation classes 4d 2d}
\cF_{R}^{(4d)} \to \cF_{R}^{(2d)} - \frac{\kappa}{2} \, t_\fg\;.
\ee
Integrating the twisted anomaly polynomial $I_{6}^{\text{twisted}}$ over $\Sigma_{\fg}$ and comparing to \eqref{I4} we read off the 2d anomaly coefficients. Assuming $9k_{RRF_{i}}=k_{F_{i}}=0$ for all flavor symmetries $F_{i}$,\footnote{This assumption is valid for many SCFTs and in even more theories to leading order in $N$, see \cite{Benini:2015bwz} for a number of examples.} one can show that the 2d trial central charge is extremized by the UV R-symmetry (see \cite{Benini:2015bwz} for details) and thus the IR and UV superconformal R-symmetries coincide and one finds the relations
\equ{
k_{RR}=(\fg-1)\,k_{RRR}\,, \qquad k=(\fg-1)\,k_{R}\,.
}
Using \eqref{R-conf anom} this can be written as
\be
\label{universalmat}
\mat{ c_r \\ c_l } = \frac{16}3\, (\fg-1) \mat{ 5 & -3 \\ 2 & 0} \mat{a_{4d} \\ c_{4d}} \;.
\ee
In the large $N$ limit this becomes
\equ{
\label{4d02univ}
c_{r}\simeq c_{l}\simeq \frac{32}{3}\,(\fg-1)\,a_{4d}\,.
}
Once again,  a positive 2d central charge requires a negatively curved Riemann surface. We derive this universal relation from holography in Section~\ref{sec:N=2 supergravity}. 

\subsubsection{$\cN=2$}
\label{sec:N=2 on Sigma}

For $\cN=2$ SCFTs there are more possibilities for twisted compactifications. The R-symmetry group is $SU(2)\times U(1)$, and we denote the Cartan generators with $R_{3}$ and $R_{0}$, respectively. A partial topological twist along $R_{3}$ preserves 2d $\cN=(2,2)$ supersymmetry, while a twist along $R_{0}$ preserves $\cN=(0,4)$ supersymmetry. Following the terminology in \cite{Kapustin:2006hi} we refer to these as the $\alpha$- and $\beta$-twist, respectively.  The twist performed for the $\cN=1$ theory in Section~\ref{N=1 Sigma} above is a particular combination of these, since when one considers the $\mathcal{N}=2$ as a particular example of an $\cN=1$ SCFT, the $\cN=1$ superconformal R-symmetry is given by the linear combination
\equ{\label{N=1 R N=2}
R_{\mathcal N=1}=\frac13 \,R_{0}+\frac43 \, R_{3}\,.
} 
We first discuss the $\alpha$-twist. Since this corresponds to a twist along the Cartan of $SU(2)$, it breaks the R-symmetry to $U(1)_{R_{3}}\times U(1)_{R_{0}}$,  with $R_{3}$ becoming the vector R-symmetry and $R_{0}$ becoming the axial R-symmetry of the $\cN=(2,2)$ theory (see, e.g., Appendix F in \cite{Putrov:2015jpa} for details). The anomaly polynomial of the 4d theory is given by
\eqs{
\label{4D anomaly form N=2}
I_{6} &= \frac{k_{R_{0}R_{0}R_{0}}}{6} \, c_{1}(\cF_{R_{0}})^{3}  +\frac{k_{R_{0}R_{3}R_{3}}}{2}\, c_{1}(\cF_{R_{0}})\, c_{1}(\cF_{R_{3}})^{2}  -\frac{k_{R_{0}}}{24} \, c_{1}(\cF_{R_{0}})\,p_{1}(\cT_{4}) \;,
}
where we have used the fact that any trace with an odd number of $R_{3}$ vanishes.  The $\alpha$-twist amounts to the shifts:
\be
\cF_{R_{3}}^{(4d)} \to \cF_{r}^{(2d)}+ \cF_{l}^{(2d)} - \frac{\kappa}{2} \, t_\fg \;,\qquad\qquad \cF_{R_{0}}^{(4d)} \to \cF_{r}^{(2d)} -\cF_{l}^{(2d)} \;.
\ee
Performing the twist in the anomaly polynomial  \eqref{4D anomaly form N=2} and  integrating over $\Sigma_{\fg}$ leads to the 2d 't Hooft anomalies
\equ{
k_{rr}=2(\fg-1)\,k_{R_{0}R_{3}R_{3}}\,, \qquad k=0\,.
}
Using the relations\footnote{These follow from \eqref{R-conf anom} and \eqref{N=1 R N=2}, and using the fact that any trace with an odd number of $R_{3}$ vanishes. } 
\equ{
k_{R_{0}R_{0}R_{0}}=k_{R_{0}}=48\,(a_{4d}-c_{4d})\,,\qquad k_{R_{0}R_{3}R_{3}}=2\,(2a_{4d}-c_{4d})\,,
}
which are valid for any 4d $\cN=2$ SCFT, one obtains the 2d central charge 
\eqs{\label{4d2d2,2} 
\text{2d $\cN=(2,2)$:}&&c_{l}=c_{r}=3k_{rr}=12\,(\fg-1)\,(2a_{4d}-c_{4d})\,.
}
It is interesting to note that the linear combination of 4d conformal anomalies, $4(2a_{4d}-c_{4d})$, appears in the Shapere-Tachikawa formula \cite{Shapere:2008zf}
\begin{equation}
4(2a_{4d}-c_{4d}) = \sum_{i=1}^{r}(2D(O_i)-1)\;.
\end{equation}
Here $r$ is the complex dimension of the Coulomb branch and $D(O_i)$ are the conformal dimensions of the operators $O_i$ which parametrize it. Therefore, one can rewrite the 2d conformal anomalies in \eqref{4d2d2,2} as a sum of dimensions of Coulomb branch operators, $c_{l}=c_{r}= 3(\fg-1)\sum_{i=1}^{r}(2D(O_i)-1)$. This clearly suggests a relation between the 2d $(2,2)$ SCFT in the IR  and the Coulomb branch of the 4d $\cN=2$ SCFT in the UV. It would be nice to understand this relation more precisely. Let us also note that if the 4d $\cN=2$ SCFT admits a Lagrangian description in terms of a vector multiplet, with gauge group $G$ of dimension $d_G$, one can show that the 2d central charges can be written as $c_{l}=c_{r}=3\,(\fg-1)\,d_{G}$.

If the 4d $\mathcal{N}=2$ SCFT admits a large $N$ limit \eqref{4d2d2,2} becomes
\equ{\label{4d2d2,2largeN}
c_{l}=c_{r}\simeq12\,(\fg-1)\,a_{4d}\,.
} 
We will derive this holographically in Section~\ref{sec:5d N=4 supergravity}. Finally we note that for $\fg=0$ the two-dimensional central charge in \eqref{4d2d2,2largeN} is negative and has the same numerical value as the central charge of the chiral algebra associated to the 4d $\cN=2$ SCFT following the procedure in \cite{Beem:2013sza}. 

We now discuss the $\beta$-twist, which amounts to the following shift of the $U(1)$ R-symmetry bundle
\be
\cF_{R_{0}}^{(4d)} \to \cF_{R}^{(2d)}- \frac{\kappa}{2} \, t_\fg \;.
\ee
The $SU(2)$ R-symmetry of the 4d theory is untouched and becomes an $SU(2)_{R}$ symmetry of the 2d $\cN=(0,4)$ theory. The central charges of the theory can be computed from an $\cN=(0,2)$ subalgebra, whose $U(1)$ R-symmetry is generated by $T=2T_{3}$, with $T_{3}$ the Cartan of $SU(2)_{R}$. This gives
\eqs{\label{clcr2004B}
\text{2d $\cN=(0,4)$:}&&\begin{pmatrix}
c_{r}\\
c_{l}
\end{pmatrix}\,=24\,(\fg-1)\begin{pmatrix}
2& -1\\
0&1
\end{pmatrix}  \begin{pmatrix}
a_{4d}\\
c_{4d}
\end{pmatrix}\,.
}
We note that the $\beta$-twist has been considered also in \cite{Putrov:2015jpa} and \cite{Cecotti:2015lab}. The expression in \eqref{clcr2004B} is the same as the one discussed in Appendix A of \cite{Cecotti:2015lab} after setting $\alpha=1$ in that paper.

For 4d $\mathcal{N}=2$ SCFTs with a Lagrangian description it is easy to check that for the $\alpha$-twist $c_r=c_{l}$ is an integer multiple of $3$ and for the $\beta$-twist $c_r$ is an integer multiple of $6$, as expected on general grounds from the (small) $\cN=4$ superconformal algebra, see for example \cite{Schwimmer:1986mf}. It is natural to conjecture that the IR 2d $\cN=(2,2)$ SCFT for the $\alpha$-twist can be described in terms of a nonlinear $\sigma$-model on a Calabi-Yau target space. It is certainly desirable to find explicitly such a description. 

As an illustration of our formulas we now consider the well-known ``rank 1'' 4d $\cN=2$ SCFTs discussed, for example, around Table 1 of \cite{Beem:2014zpa}. These SCFTs are distinguished by having a one-dimensional Coulomb branch of vacua and we summarize the results for the two-dimensional central charges for the three universal twists we have studied so far, namely the $\alpha$- and $\beta$-twists discussed above as well as the universal $(0,2)$ twist discussed in Section \ref{N=1 Sigma}, in Table \ref{tab:table1}. Some comments are in order. First, note that for all theories in Table \ref{tab:table1}, except the $H_0$ and $H_1$ theories, $c_r^{(\alpha)} = c_l^{(\alpha)}$ is an integer multiple of $3$ and $c_r^{(\beta)}$ is an integer multiple of $6$, as should be the case. Theories $H_0$ and $H_1$ should be treated more carefully in view of the quantization conditions on R-charges discussed around \eqref{QC}. The operator with lowest R-charge in the $H_0$ theory has the value $r_{H_0} = 2/5$ and the one in the $H_1$ theory has $r_{H_1} = 2/3$. It follows from the quantization condition \eqref{QC} that for the $H_0$ theory the $\alpha$- and $\beta$-twist are well-defined only when $(\fg-1)$ is an integer multiple of $5$. For the $H_1$ theory, by the same analysis, $(\fg-1)$ should be an integer multiple of $3$. These constraints then ensure that also for these two SCFTs,  $c_r^{(\alpha)} = c_l^{(\alpha)}$ is an integer multiple of $3$ and $c_r^{(\beta)}$ is an integer multiple of $6$. It is also curious to note the following relations between the 2d conformal anomalies and the dual Coxeter number of the flavor symmetry group of the four-dimensional theory:\footnote{These relations are of course not obeyed for the $H_0$ theory in Table \ref{tab:table1} since this theory does not have a flavor symmetry.}
\begin{eqnarray}
\begin{split}
c_r^{(\alpha)} &= c_l^{(\alpha)}= (h^{\vee}+3)(\fg-1)\;, \quad c_r^{(\beta)} = 2(h^{\vee}+3)(\fg-1)\;, \quad c_l^{(\beta)} = 4(h^{\vee}+1)(\fg-1)\;, \\
c_r^{(0,2)} &= \frac{2}{9}\,(3h^{\vee}+13)(\fg-1)\;, \qquad c_l^{(0,2)} = \frac{4}{9}\,(3h^{\vee}+5)(\fg-1)\;.
\end{split}
\end{eqnarray}
%

\begin{table}[h!]
  \centering
    \begin{tabular}{|c||c|c|c|c|c|c|c|} \hline
     & $H_0$ & $H_1$ & $H_2$ & $D_4$ & $E_6$ & $E_7$ & $E_8$ \\
    \hline \hline
    $G_F$ & - & $SU(2)$ & $SU(3)$ & $SO(8)$ & $E_6$ & $E_7$ & $E_8$ \\
    \hline
    $h^{\vee}$ & - & $2$ & $3$ & $6$ & $12$ & $18$ & $30$ \\
    \hline
    $c_{4d}$ & $\frac{11}{30}$ & $\frac{1}{2}$ & $\frac{2}{3}$ & $\frac{7}{6}$ & $\frac{13}{6}$ & $\frac{19}{6}$ & $\frac{31}{6}$ \\
    \hline
    $a_{4d}$ & $\frac{43}{120}$ & $\frac{11}{24}$ & $\frac{7}{12}$ & $\frac{23}{24}$ & $\frac{41}{24}$ & $\frac{59}{24}$ & $\frac{95}{24}$ \\
      \hline 
    $c_r^{(\alpha)}=c_l^{(\alpha)}$ & $\frac{21}{5} (\fg-1)$ & $5 (\fg-1)$ & $6 (\fg-1)$ & $9 (\fg-1)$ & $15(\fg-1)$ & $21(\fg-1)$ & $33 (\fg-1)$ \\
    \hline
    $c_r^{(\beta)}$ & $\frac{42}{5} (\fg-1)$ & $10 (\fg-1)$ & $12 (\fg-1)$ & $18 (\fg-1)$ & $30(\fg-1)$ & $42(\fg-1)$ & $66 (\fg-1)$ \\
      \hline
    $c_l^{(\beta)}$ & $\frac{44}{5} (\fg-1)$ & $12 (\fg-1)$ & $16 (\fg-1)$ & $28 (\fg-1)$ & $52(\fg-1)$ & $76(\fg-1)$ & $124 (\fg-1)$ \\
       \hline
    $c_r^{(0,2)}$ & $\frac{166}{45} (\fg-1)$ & $\frac{38}{9} (\fg-1)$ & $\frac{44}{9} (\fg-1)$ & $\frac{62}{9} (\fg-1)$ & $\frac{98}{9}(\fg-1)$ & $\frac{134}{9}(\fg-1)$ & $\frac{206}{9} (\fg-1)$ \\
    \hline
     $c_l^{(0,2)}$ & $\frac{172}{45} (\fg-1)$ & $\frac{44}{9} (\fg-1)$ & $\frac{56}{9} (\fg-1)$ & $\frac{92}{9} (\fg-1)$ & $\frac{164}{9}(\fg-1)$ & $\frac{236}{9}(\fg-1)$ & $\frac{380}{9} (\fg-1)$ \\
     \hline
  \end{tabular}
  \caption{Flavor symmetries, dual Coxeter numbers, and 4d and 2d conformal anomalies for ``rank 1'' SCFTs.}
  \label{tab:table1}
\end{table}
We note, as a consistency check, that the relation \eqref{clcr2022} for compactifications from 6d to 2d can be obtained by first compactifying the 6d theory to 4d by the MN $\cN=2$ twist \eqref{matMN2} on $\Sigma_{1}$, and subsequently compactifying to 2d by the $\alpha$-twist \eqref{4d2d2,2}  on $\Sigma_{2}$. Similarly, the 6d to 2d flow relation in \eqref{c2dn=2004} for the case when the K\"ahler manifold is a product of Riemann surfaces can be obtained by composing the MN $\cN=2$ twist relation \eqref{matMN2} with  the $\beta$-twist result in \eqref{clcr2004B}. 

\paragraph{Compactification on $M_3$.}

One may consider the twisted compactification of a 4d $\cN=2$ SCFT on a three-manifold with holonomy $SO(3)\simeq SU(2)$ by turning a background gauge field for the $SU(2)_R$ symmetry. This twist preserves two real supercharges and thus leads to a 1d supersymmetric quantum mechanics. Since we do not have anomalies at our disposal in 1d, we cannot apply the procedure above and we are not able to say much about the properties of the resulting 1d theory. However, for $M_3=\Bbb H^3$ the RG flow across dimensions can be constructed holographically (see Section~\ref{sec:N=2 supergravity}), which corresponds to a magnetically charged BPS black hole with an AdS$_2$ near-horizon geometry in locally AdS$_5$. This suggests that the quantum mechanical theory in the IR has an emergent conformal symmetry, at least at large $N$.  We comment further on this model in Section~\ref{sec:N=2 supergravity}.

\subsubsection{Comments on $\cN=3$ and $\cN=4$ SCFTs}
\label{sec:N=34 on Sigma}

It has been recently observed \cite{Aharony:2015oyb,Garcia-Etxebarria:2015wns,Aharony:2016kai} that four-dimensional theories with $\cN = 3$ supersymmetry, and no $\cN=4$ supersymmetry, can exist at strong coupling. These have a $U(3)$ R-symmetry, no flavor symmetries, and it can be shown  that  $a_{4d}=c_{4d}$ \cite{Aharony:2015oyb}. The twisted compactification of these theories on a Riemann surface was considered in \cite{Amariti:2017cyd}.  Since the R-symmetry group is contained in that of  $\cN=4$ SYM, and the ranks are the same, the possible twists of $\cN=3$ theories are contained in those possible for $\cN=4$ SYM, studied in \cite{Benini:2013cda}. Viewing an $\cN=3$ theory as an $\cN=1$ or $\cN=2$ theory with additional flavor symmetry, one can apply the twists discussed in Sections \ref{N=1 Sigma} and \ref{sec:N=2 on Sigma}.  In particular, the $\alpha$- and $\beta$-twists of the $\cN=2$ theory applied to an $\cN=3$ theory leads to a 2d theory with $\cN=(2,2)$ and $\cN=(2,4)$ supersymmetry, respectively. Since only the second twist preserves half the amount of supercharges, it is a genuine universal twist of the $\cN=3$ theory, in the sense used here. 
The resulting central charges are given by
\eqs{\label{N=3to2,4}
\text{2d $\cN=(2,4)$:}&&c_{l}=c_{r}=24\, (\fg-1)\, a_{4d}\,,
}
which is obtained by simply setting $a_{4d}=c_{4d}$ in \eqref{clcr2004B}.\footnote{Of course, one may still apply the $\alpha$-twist to an $\cN=3$ theory, and the corresponding central charges are given by $c_{l}=c_{r}=12\, (\fg-1)\, a_{4d}$, which follows by setting $a_{4d}=c_{4d}$ in  \eqref{4d2d2,2}.} Another twist preserving six supercharges arises from taking $U(3)=SU(3)\times U(1)$ and twisting along the $U(1)$. This leads to  a 2d theory with $\cN=(0,6)$ supersymmetry, $U(3)$ R-symmetry, and central charges
\eqs{\label{N=3to0,6} 
\text{2d $\cN=(0,6)$:}&&c_{l}=c_{r}=\frac{96}{5}\, (\fg-1)\, a_{4d}\,.
}
Note that one always obtains a 2d theory with no gravitational anomaly, $c_{r}-c_{l}=0$, as a consequence of  the absence of gravitational anomalies in 4d $\cN=3$ theories. For a discussion of other twists of $\cN=3$ theories see \cite{Amariti:2017cyd}. We refer the reader to  \cite{Benini:2013cda} for general twists of $\cN=4$ SYM. 
 
\subsection{SCFTs in odd dimensions}
\label{sec:Comments on 3d and 5d SCFTs}

Based on the results  for flows between even-dimensional SCFTs described above, it is natural to wonder if analogous universal relations exist when at least one of the fixed points in the RG flow is an odd-dimensional SCFT. Since there are no 't Hooft and conformal anomalies in odd dimensions, and in view of the $F$-theorem \cite{Jafferis:2011zi}, one may look for universal relations involving an appropriately defined partition function, or free energy, of the odd-dimensional SCFT. 

The simplest case for which this can be explored by pure field-theoretic methods is the case of three-dimensional $\cN=2$ theories placed on a Riemann surface $\Sigma_{\fg}$. The two quantities we wish to compare in this case are the $S^{3}$ partition function of the theory before compactification and the partition function on  $S^{1}\times \Sigma_{\fg}$ with a partial topological twist on $\Sigma_{\fg}$. The former, which we denote by  $Z_{S^{3}}(\Delta_{I})$ was computed by supersymmetric localization in \cite{Kapustin:2009kz,Jafferis:2010un} and is a function of trial R-charges $\Delta_{I}$ for the theory on $S^{3}$. The latter, which we denote by $Z(y_{I},\mathfrak n_{I})$, was computed by localization in  \cite{Benini:2015noa,Benini:2016hjo,Closset:2016arn}. It is a function of background magnetic fluxes $\mathfrak n_{I}$ specifying the topological twist and flavor fugacities $y_{I}$ and can be interpreted as a twisted index of the 3d theory or a Witten index of the 1d low-energy theory. 

Progress in understanding the relation between these quantities was  made recently in    \cite{Benini:2015eyy,Benini:2016hjo,Hosseini:2016tor,ABCMZ,BMP}, which we summarize next. Supersymmetric localization shows that both partition functions  localize to a  matrix model on the Coulomb branch of the theory.  Although the resulting matrix models  appear to be quite different at finite $N$, it has been observed that for a large class of Chern-Simons matter theories at large $N$ the partition functions are in fact intimately related as follows:\footnote{This was first shown for $\Sigma_{\fg}=S^{2}$ in  \cite{Hosseini:2016tor}, for a generic genus in the case of ABJM theory in \cite{Benini:2016hjo}, and for a larger class of quiver theories and  generic Riemann surface in  \cite{ABCMZ}.}
\equ{\label{indexS3}
\text{Re} \log Z(y_{I},\mathfrak n_{I})= (\fg-1)\left[F_{S^{3}}\(\frac{\Delta_{I}}{\pi}\)+\sum_{I}\(\frac{\mathfrak n_{I}}{1-\fg}-\frac{\Delta_{I}}{\pi}\)\frac{\pi}{2}\frac{\partial }{\partial \Delta_{I}}F_{S^{3}}\(\frac{\Delta_{I}}{\pi}\)\right]\;,
}
where one makes the identification $y_{I}=e^{i\Delta_{I}}$, and subject to the constraints $\sum_{I}\mathfrak n_{I}=2(1-\fg)$ and $\sum_{I}\Delta_{I}=2\pi$.\footnote{The constraint on  $\sum_{I}\Delta_{I}$ is rather subtle and requires a detailed large $N$ analysis of the matrix model. See \cite{ABCMZ} for a more detailed discussion.} The observation made in  \cite{ABCMZ} is that  many 3d $\cN=2$ theories admit a universal topological twist,\footnote{This is consistent only if the exact R-charges in the SCFT are rational. Although this is not a problem for many 3d $\cN=2$ theories, there are an infinite number of examples for which this issue arises. See \cite{ABCMZ,BMP} for a more detailed discussion.} which amounts to setting the flux parameters $\mathfrak n_{I}$ to be proportional to the exact R-charges $\bar \Delta_{I}$ of the theory on $S^{3}$. Imposing this in \eqref{indexS3} and  denoting $F_{S^{1}\times \Sigma_{\fg}}(\bar \Delta_{I})\equiv -\text{Re} \log Z(\bar \Delta_{I})$, the end result is the simple large $N$ universal relation
\equ{\label{F3d1d}
F_{S^{1}\times \Sigma_{\fg}}(\bar \Delta_{I})=(1-\fg)\,F_{S^{3}}(\bar \Delta_{I}/\pi)\,.
}
This is the first example we encounter of a universal relation of the form \eqref{eqFintro}.  It would be interesting to determine the exact role of subleading orders in $N$ in this relation.  As we discuss in Section~\ref{sec:4d supergravity}, this twisted compactification is described holographically by a magnetically charged black hole in AdS$_{4}$, whose entropy at large $N$ is computed by $F_{S^{1}\times \Sigma_{\fg}}$. This gives further motivation for studying subleading corrections to this quantity. 

Let us comment  on 3d SCFTs with $\mathcal{N}>2$. Clearly, all such theories can be viewed as  $\mathcal{N}=2$ theories and one can readily apply the results discussed above. However, as in the case of the $\alpha$- and $\beta$-twists of 4d theories discussed in Section \ref{sec:N=2 on Sigma}, one may wonder if twisted compactifications with enhanced supersymmetry  exhibit any universal properties. For $\mathcal{N}=3$ SCFTs the R-symmetry is $SO(3)$ and thus the only topological twist available is the universal one. We note that if the $\mathcal{N}=3$ theory at hand admits a Lagrangian description (and perhaps even more generally), the R-charges are rational and thus the universal twist can always be performed for some appropriate value of the genus, $\fg>1$. For 3d $\cN=4$ SCFTs the situation is more interesting.\footnote{We refrain from discussing SCFTs with $\mathcal{N}>4$ here.} These theories have $SU(2)_C\times SU(2)_H$ R-symmetry and thus admit more general topological twists on $\Sigma_{\fg}$. The $\mathcal{N}=2$ universal twist preserves two real supercharges and amounts to turning on a magnetic flux along the Cartan generator of the diagonal $SU(2)$ subgroup of the R-symmetry group. There are, however, two other twists which preserve four real supercharges and are obtained by turning on a magnetic flux along the Cartan generator of either $SU(2)_C$ or $SU(2)_H$. These twisted compactifications were recently studied in \cite{Gaiotto:2016wcv}. One can then study the topologically twisted index for both of these twists using the results of \cite{Benini:2015noa,Benini:2016hjo,Closset:2016arn}. In order to make a connection with the universal relations derived above, and ultimately with holography, we are interested in the large $N$ limit of these indices. It turns out, however, that the topologically twisted index is trivial to leading order in $N$ for both topological twists, i.e. the corresponding free energy vanishes.\footnote{We are grateful to Alberto Zaffaroni for informing us of this result.} It would certainly be interesting to explore these two twisted compactifications further and understand whether the topologically twisted indices obey any type of universal relation at finite $N$.
 
Let us briefly comment  on the case $d=5$. The only superconformal algebra in five dimensions is $F(4)$, with eight supercharges and an $SU(2)_{R}$ R-symmetry.  The twisted compactification of such SCFTs on a Riemann surface $\Sigma_{\fg}$ by twisting along the Cartan of $SU(2)_{R}$ leads to a 3d SCFT with  $\cN=2$ supersymmetry. The twisted compactification on a three-manifold $M_{3}$ by a nonabelian twist along the full $SU(2)_{R}$ leads to a  2d SCFT with $\cN=(1,1)$ supersymmetry. Finally, one may also consider the compactification on a K\"ahler four-manifold, leading to supersymmetric quantum mechanical theories. The holographic description of these RG flows across dimensions was considered in \cite{Nunez:2001pt,Naka:2002jz}, which we discuss in Section \ref{sec:6d  supergravity}. While the partition function of 5d SCFTs on various five-manifolds has been studied extensively in the literature, much less is known for partially twisted compactifications.\footnote{See however \cite{Fukuda:2012jr} for a localization calculation of 5d theories on $S^3\times \Sigma_{\fg}$ with a topological twist on $\Sigma_{\fg}$.} In view of the simple holographic relations between the UV and IR free energies (or conformal anomalies in the case of $M_3$) that we uncover in Section \ref{sec:6d  supergravity}, it would be interesting to explore this further in field theory. 

Finally, let us point out that for RG flows in which $d$ is even and $p$ is odd (or vice versa) it is tempting to look for universal relations among free energies and conformal anomalies. While we have not studied this in field theory, and are not aware of any discussion in the literature,\footnote{See \cite{Giombi:2014xxa} for a related but distinct discussion on the connection between conformal anomalies and sphere free energies for theories in different dimensions.} the holographic analysis in Section \ref{sec:Gauged Supergravity} provides evidence that such universal relations exist. This  certainly deserves further study.

\subsection{Comments on Hofman-Maldacena bounds}
\label{sec:Comments on Hofman-Maldacena bounds}

The Hofman-Maldacena (HM) bounds are bounds on the ratio $a_{4d}/c_{4d}$ in four-dimensional CFTs derived from energy positivity constraints. These were first proposed in \cite{Hofman:2008ar} (see also \cite{Shapere:2008zf,Buchel:2009sk}), and were recently proven in \cite{Hofman:2016awc} using conformal bootstrap methods. For supersymmetric theories these bounds read:  
\eqss{
\cN=1:\quad \frac12\leq\frac{a_{4d}}{c_{4d}}\leq \frac32\,, \qquad \qquad \cN=2:\quad \frac12\leq\frac{a_{4d}}{c_{4d}}\leq\frac54\,.
\label{eq:HM}
}
It is interesting to study how  various values of $a_{4d}/c_{4d}$ are mapped to six and two dimensions by the RG flows across dimensions discussed above.

\begin{figure}[]
\begin{center}
\includegraphics{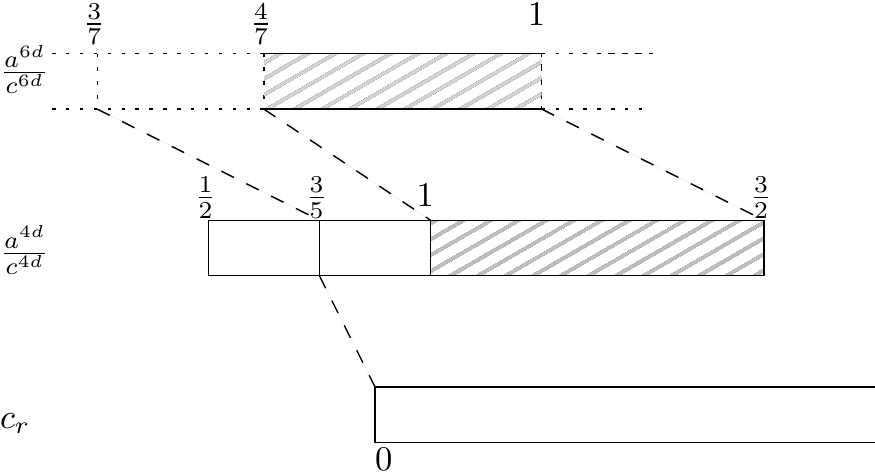}
\caption{Mapping of various values of $a/c$ under  \eqref{ac4dn=120} from 6d $\cN=(2,0)$ to a 4d $\cN=1$ theory, as well as the flow from 4d to 2d obeying \eqref{universalmat} and  preserving $\cN=(0,2)$ supersymmetry. Positivity of $c_{r}$ in 2d is mapped into the lower bound $3/7$ in 6d.}
\label{univ_twist_bounds1}
\end{center}
\end{figure}

Consider first 4d $\cN=1$ theories obtained by the twisted compactification of  6d $\cN=(2,0)$ theories on a Riemann surface. The four- and six-dimensional central charges are related by  \eqref{ac4dn=120}. One can further compactify  to 2d by the universal twist preserving 2d $\cN=(0,2)$ supersymmetry. The two- and four-dimensional central charges are then related by \eqref{universalmat}. In Figure~\ref{univ_twist_bounds1} we show how various values are mapped across dimensions. We observe that  6d  ADE theories  (shaded region) always lead to 4d $\cN=1$ theories satisfying the HM bounds, with the upper HM bound in 4d saturated when $a_{6d}/c_{6d}=1$. Since the value $\frac{a_{4d}}{c_{4d}}=\frac{3}{2}$ is associated to a free 4d $\mathcal{N}=1$ vector multiplet it is natural to conjecture that six-dimensional $\cN=(2,0)$ SCFTs with $a_{6d}=c_{6d}$ flow to a free gauge theory upon this twisted compactification.  Upon further compactification to 2d we note that the positivity constraint $c_{r}\geq0$ imposed by 2d unitarity requires $a_{6d}/c_{6d}\geq 3/7$.\footnote{It is curious to note that this lower bound on $a_{6d}/c_{6d}$ appeared also in \cite{Zhou:2016kcz} in a different context} Notice also that the two dimensional unitarity bound, $c_r>0$, translates into the bound $a_{4d}/c_{4d}>3/5$ for 4d $\mathcal{N}=1$ SCFTs. A free $\mathcal{N}=1$ chiral multiplet has the ratio $a_{4d}/c_{4d}=1/2$ but we are not aware of any other 4d unitary $\mathcal{N}=1$ SCFTs which have a ratio of conformal anomalies in the range $1/2 \leq a_{4d}/c_{4d} <3/5$. A natural question arising from this observation is whether one can use two-dimensional unitarity in combination with the universal RG flows from two to four dimensions to derive a stronger bound on the ratio $a_{4d}/c_{4d}$ than the one in \eqref{eq:HM}.

\begin{figure}[h!]
\begin{center}
\includegraphics[]{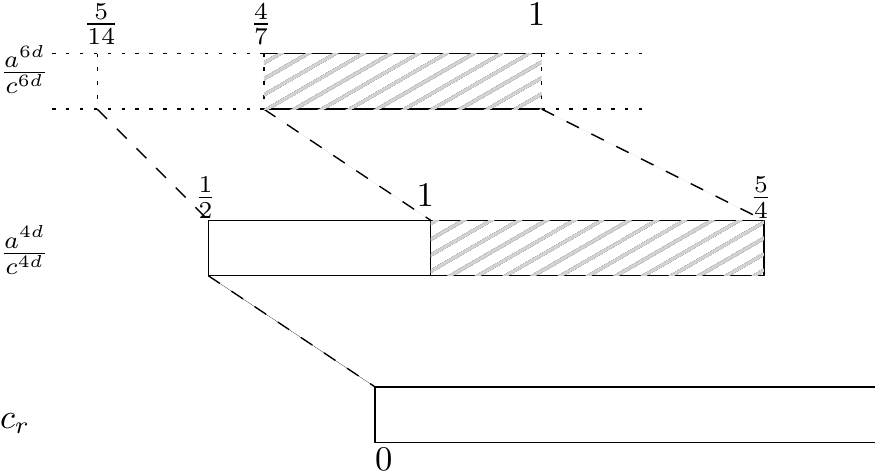}
\caption{Mapping of various values of $a/c$ under  \eqref{matMN2} from 6d $\cN=(2,0)$ to a 4d $\cN=2$ theory. The  map to 2d, either by  the $\alpha$-twist \eqref{4d2d2,2} or the $\beta$-twist \eqref{clcr2004B}, relates the lower HM bound in 4d to the 2d unitarity bound $c_{r}\geq0$ ($c_{l}\geq0$ is automatically satisfied). }
\label{univ_twist_bounds2}
\end{center}
\end{figure}

Consider now 4d $\cN=2$ theories obtained by twisted compactification of  6d $\cN=(2,0)$ theories on a Riemann surface; the central charges are related by \eqref{matMN2}. Once again, 6d ADE theories lead to 4d theories satisfying the HM bounds, with the upper HM bound in 4d saturated when $a_{6d}/c_{6d}=1$. Again, this naturally suggests that a 6d $\cN=(2,0)$ theory with $a_{6d}=c_{6d}$ leads to the theory of a free $\mathcal{N}=2$ vector multiplet upon this twisted compactification. One may further reduce to 2d, either by the $\alpha$-twist \eqref{4d2d2,2}, or the $\beta$-twist \eqref{clcr2004B}. In both cases the lower HM bound in 4d is mapped to $c_{r}=0$ in 2d, as shown in Figure~\ref{univ_twist_bounds2}. This is very intriguing and suggests a relation between unitarity in two dimensions and the bounds in \eqref{eq:HM}. As noted in Section \ref{subsec:Weyl}, the 6d $\cN=(2,0)$  ADE theories\footnote{These are the only currently known $\cN=(2,0)$ SCFTs in six dimensions.} have conformal anomaly coefficients  obeying $4/7< a_{6d}/c_{6d}<1$. Upon the $\mathcal{N}=2$ twisted compactification on $\Sigma_{\fg}$ and using \eqref{matMN2} this implies that the 4d class $\mathcal{S}$ SCFT has central charges  obeying  $a_{4d}/c_{4d}>1$. Note also that this class of 4d SCFTs arise from M5-branes wrapping $\Sigma_{\fg}$ and have well-known holographic dual descriptions \cite{Maldacena:2000mw,Gaiotto:2009gz}. This observation is in conflict with the recent results in \cite{Cheung:2016wjt} where the authors argued for a positivity bound on the Gauss-Bonnet coupling in higher-curvature gravitational theories. In particular, the results in  \cite{Cheung:2016wjt}  imply that for CFTs with a weakly coupled holographic dual, such as the theories at hand, one should find $a_{4d}/c_{4d}<1$. It would be interesting to understand the reasons for this inconsistency.

%

\section{Gauged supergravity}
\label{sec:Gauged Supergravity}

We now turn to the holographic description of the universal RG flows described in Section~\ref{sec:Field theory}. As we have argued, when the topological twist involves only the stress-energy tensor multiplet the RG flow is universal and is not sensitive to the details of the particular theory being compactified. General results in superconformal representation theory show (see for example \cite{Buican:2016hpb,Cordova:2016emh} for a recent account) that the stress-energy multiplet for SCFTs in $d=3,4,5,6$ has an operator spectrum exactly dual to the fields in the gravity multiplet of a gauged supergravity in one dimension higher. This suggests that the holographic description of universal RG flows across dimensions should be captured entirely by the dynamics of the gravity multiplet in the appropriate gauged supergravity. We refer to a supergravity theory containing only the gravity multiplet as {\it minimal} gauged supergravity.\footnote{This might not be standard terminology but we employ it here to emphasize that there are no other multiplets apart from the gravity multiplet.}  Motivated by the analysis of Section~\ref{sec:Field theory}, in this section we discuss supersymmetric solutions of minimal gauged supergravities in 4, 5, 6, and 7 dimensions describing universal RG flows. 

We impose the following conditions on the supergravity solutions. At asymptotic infinity the metric should approach an asymptotically locally AdS$_{d+1}$ background, with an $\Bbb R^{p}\times M_{d-p}$ boundary. In the interior of the geometry there should be another asymptotic region, where the metric  approaches AdS$_{p+1}\times M_{d-p}$. This is captured by an Ansatz of the form
\equ{
ds^{2}_{d+1}=e^{2f(r)}\,ds^{2}_{\Bbb R^{p+1}}+e^{2g(r)}\, ds^{2}_{M_{d-p}}\,,
}
with the following asymptotics:
\eqs{\label{UVfp}
r\to \infty:&& e^{2f(r)}&=\frac{1}{r^{2}}\,,&& e^{2g(r)}=\frac{1}{r^{2}}\,, \\ \label{IRfp}
r\to 0:&& e^{2f(r)}&=\frac{e^{2f_{0}}}{r^{2}}\,, && e^{2g(r)}=e^{2g_{0}}\,, 
}
where $r$ is the holographic direction, with the UV corresponding to $r\to\infty$ and the IR to $r\to0$, and $f_{0},g_{0}$ are constants determined by the supergravity BPS equations. In addition, there must be a nontrivial graviphoton gauge field $A_{\mu}^{R}$ , proportional to the spin connection on $M_{d-p}$. This gauge field flux, which is the holographic manifestation of the topological twist, ensures that the solution preserves a certain amount of supersymmetry.  In theories with more than 8 supercharges, there are also scalars in the gravity multiplet which generically also acquire a nontrivial radial profile as a function of $r$. The entire supergravity solution can be viewed as a  $(p-1)$-dimensional BPS black brane in AdS$_{d+1}$ carrying a magnetic charge under the graviphoton $A_{\mu}^{R}$. We emphasize that the magnetic charge of the black brane is fixed to a unique value by supersymmetry.\footnote{Generalizations including charges under additional vector multiplets, preserving the same amount of supersymmetry are possible. However, these solutions will not be universal, as argued in the Introduction. } Since for the purposes of our discussion it suffices to focus on the AdS$_{p+1}$ and AdS$_{d+1}$ asymptotics of the solution, we will not present the full interpolating domain wall  explicitly.\footnote{The full radial dependence of the metric functions and  possible scalar fields  can be found analytically in some examples but in general one has to resort to numerical integration of the BPS equations.}

The careful reader may have noticed a sleight of hand in our discussion. When one performs the topologically twisted reduction in the field theory there is no guarantee that the IR dynamics of the QFT will be governed by an interacting SCFT with a weakly coupled holographic dual. We have allowed for this possibility in our discussion below, by allowing the IR region of the gravitational domain wall to have a metric different from AdS$_{p+1}$ and studying the BPS equations in this more general setup. All such solutions, however,  turn out to be singular, suggesting that the corresponding twisted compactifications  have some ``pathological'' behavior in the IR. For instance, the IR theory could become free, have a non-normalizable vacuum, or accidental global symmetries.

We should emphasize that there is a vast literature on constructing domain wall solutions in gauged supergravity. An important vantage point on these solutions was offered by the work of Maldacena and N\'u\~nez \cite{Maldacena:2000mw}, where twisted compactifications of  6d $\cN=(2,0)$ theories and 4d $\cN=4$ SYM on Riemann surfaces were studied holographically, and  shown to correspond to supergravity backgrounds of the kind discussed above. Our main goal in this section is not to find new supergravity solutions, but rather to collect and organize various backgrounds scattered throughout the literature, and to interpret them in the context of the universal flows discussed in Section~\ref{sec:Field theory}. In particular, a crucial point in our story is that these universal black brane solutions can be embedded in ten- or eleven-dimensional supergravity in infinitely many ways (see Section~\ref{subsec:Uplifts} for a more detailed discussion).  This is precisely the statement of universality of the solutions. Different embeddings of these supergravity backgrounds in string or M-theory describe the twisted compactification of different SCFTs, but the universal relations for field theory observables discussed in Section~\ref{sec:Field theory} always hold, regardless of the details of the SCFTs. This point of view not only establishes the existence of RG flows across dimensions for a large class of SCFTs (at least to leading order in $N$), but can also be a powerful tool in counting the microstates of infinite families of black branes. This was made explicit for the case of black holes in AdS$_{4}$ in \cite{ABCMZ}, following the approach of \cite{Benini:2015eyy}. It is natural to expect that the observations made here will be useful in generalizing these results to even larger classes of black branes in different dimensions.

Before describing the solutions of interest, let us collect some expressions for the holographic evaluation of  central charges and free energies, which  will be used repeatedly below. For odd-dimensional AdS solutions the central charges of the dual even-dimensional CFTs are given by \cite{Brown:1986nw,Henningson:1998gx}:
\eqs{\label{achol}
c_{r}=c_{l}=\frac{3L_{\text{AdS}_{3}}}{2G_{N}^{(3)}}\,, \qquad a_{4d}=c_{4d}=\frac{\pi L_{\text{AdS}_{5}}^{3}}{8 G_{N}^{(5)}}\,, \qquad a_{6d}=\frac47 \, c_{6d}=\frac{3\pi^{2}L_{\text{AdS}_{7}}^{5}}{7G_{N}^{(7)}}\,.
}
Here $G_{N}^{(d+1)}$ is the Newton constant in $(d+1)$ dimensions and $L_{\text{AdS}_{d+1}}$ is the length scale associated with the given AdS vacuum. In the case of even-dimensional AdS solutions we are interested in the renormalized value of the on-shell action, which is mapped holographically to the free energy of the dual SCFT placed on a round sphere. In the case of AdS$_4$ and AdS$_6$ vacua one finds the following expressions for the sphere free energy (see for example \cite{Emparan:1999pm}): 
\equ{\label{Fgrav S3}
F_{S^{3}}=\frac{\pi L_{\text{AdS}_{4}}^{2}}{2G_{N}^{(4)}}\;, \qquad F_{S^{5}}=-\frac{\pi^{2} L_{\text{AdS}_{6}}^{4}}{3G_{N}^{(6)}}\,.
}
One should not be bothered by the minus sign in $F_{S^{5}}$ since the ``proper'' monotonically decreasing quantity under RG flow in an odd dimension $d$ is conjectured to be given by $(-1)^{(d-1)/2}\log Z_{S^d}$ \cite{Jafferis:2011zi}, where $Z_{S^d}$ is the partition function on $S^d$ and we always define $F_{S^{d}}=-\log Z_{S^d}$. 

In the case of AdS$_{2}$ vacua, describing the near-horizon geometry of BPS black holes, the main quantity of interest will be the Bekenstein-Hawking entropy of the black hole. As shown explicitly in \cite{ABCMZ} for black holes in AdS$_{4}$, the black hole entropy is intimately related to the renormalized gravitational free energy of the solution.\footnote{It would be interesting to study this relation for more general AdS black holes in various dimensions.} 

Finally, all the solutions we discuss are locally asymptotic to the AdS$_{d+1}$ vacuum of the gauged supergravity theory. The length scale of this vacuum is set by the value of the scalar potential of the theory at its AdS$_{d+1}$ critical point, $V|_{\phi_{\text{crit}}}$. This value in turn sets the cosmological constant scale, which is related to the value of the gauge coupling constant in the supergravity theory. In our conventions the radius, $L_{\text{AdS}_{d+1}}$, of AdS$_{d+1}$ is given by
\equ{\label{normalization LAdS}
V|_{\phi_{\text{crit}}}=-\frac{d\,(d-1)}{L_{\text{AdS}_{d+1}}^2}\,.
}
We choose a normalization in which $L_{\text{AdS}_{d+1}}=1$, thus fixing the value of the gauged supergravity coupling constant to a particular value, as determined by \eqref{normalization LAdS}. 

\subsection{7d  supergravity}
\label{sec:7d  supergravity}

In this section, we provide the holographic description of  RG flows from 6d $\cN=(1,0)$ and $\cN=(2,0)$ SCFTs to lower-dimensional SCFTs by the twisted compactifications described in Section~\ref{sec:6d SCFTs}. The relevant supergravity theories are the 7d $\cN=2$ gauged supergravity of \cite{Townsend:1983kk} (see also \cite{Lu:1995hm} for some details) and the 7d $\cN=4$ gauged supergravity of \cite{Pernici:1984xx}, respectively.

\subsubsection{$\cN=2$}
\label{sec:Minimal 7d gauged supergravity}

The bosonic content of this minimally supersymmetric theory is the graviton, an $SU(2)$ graviphoton, a real scalar $\lambda$, and a three-form potential $ C_{3}$. In the solutions of interest, the gauge field is excited only along the Cartan of $SU(2)$ and $ C_{3}=0$, in which case the bosonic Lagrangian reads 
\eqss{\label{Lag min 7d supergravity}
\mathcal L&=R-20\ast d\lambda \wedge d\lambda-V(\lambda)\ast 1-e^{-4\lambda}\ast  F \wedge  F \,;\\
V&=\frac{1}{2}\,m^{2}\,\(-8e^{4\lambda}-8 e^{-6\lambda}+e^{-16\lambda}\)\,,
}
where $m$ is the gauge coupling constant.  The AdS$_{7}$ vacuum of the theory corresponds to the extremum of the potential  at $\lambda=0$, where it takes the value  $V|_{\lambda=0}=-15\, m^{2}/2$. Comparing to \eqref{normalization LAdS} we  set $m=2$ in what follows to  normalize  $L_{\text{AdS}_{7}}=1$. Universal holographic RG flow solutions in this theory were studied in \cite{Passias:2015gya,Apruzzi:2015wna}.

\paragraph{AdS$_{5}$ vacua.}

We are interested in supersymmetric solutions of the form
\eqss{\label{75}
ds^{2}&=e^{2f(r)}(-dt^{2}+dr^{2}+dz_{1}^{2}+dz_{2}^{2}+dz_{3}^{2})+e^{2g(r)}\, ds^{2}_{\Sigma_{\fg>1}}\,,\\
F&=\frac{1}{8}\, \dvol_{\Sigma_{\fg>1}}\,.
}
We focus on $\fg>1$ since there are no regular solutions for $\fg=0,1$. To obtain the BPS equations we note that when the gauge field points along the Cartan of $SU(2)$, as in the Ansatz above, the minimal theory can be obtained as a subsector of the $U(1)^{2}$ truncation of the 7d $SO(5)$ maximally supersymmetric gauged supergravity.\footnote{More precisely, the  bosonic content of this truncation is two gauge fields $A_{\mu}^{(1)},A_{\mu}^{(2)}$, a three form potential $C_{\mu\nu\rho}$ and two scalar fields $\lambda_{1},\lambda_{2}$. To obtain the minimal theory one must set $A_{\mu}^{(1)}=A_{\mu}^{(2)}\equiv A_{\mu}$ and $\lambda_{1}=\lambda_{2}\equiv \lambda$. One can check, for instance, that the Lagrangian and supersymmetry variations obtained in this way are consistent with those of \cite{Liu:1999ai}.}  General RG flows in this  $U(1)^{2}$ truncation  were studied in \cite{Bah:2012dg}. We may thus borrow the BPS equations for an Ansatz of the form \eqref{75} from this reference,\footnote{These can also be obtained from the BPS equations derived in \cite{Maldacena:2000mw}.}  which read
\begin{align}
\begin{split}
e^{-f}f'&=  - 2 \,e^{-f}\, \lambda'- e^{-8\lambda}\,, \\
e^{-f}g'&= - 2\,e^{-f}\,\lambda' + \ds\frac{1}{4} \,e^{-2g-2\lambda}+ e^{-8\lambda}\,,\\
e^{-f}\lambda'&=  - \frac25\,(e^{-8\lambda} -e^{2\lambda}) +  \ds\frac{1}{40} \,e^{-2g-2\lambda} \,,
\label{BPSeqns}
\end{split}
\end{align}
where a prime denotes differentiation with respect to $r$. These equations describe the full RG flow from AdS$_{7}$ to AdS$_{5}\times \Sigma_{\fg}$. At the AdS$_{5}$ fixed point the metric functions take the form \eqref{IRfp}, with  
\equ{
e^{f_{0}}=e^{8\lambda_{0}}\,, \qquad e^{g_{0}}=\frac12 \, e^{3\lambda_{0}}\,, \qquad e^{10\lambda_{0}}=\frac34\,, 
}
where $\lambda_{0}$ is the value of the scalar field at the fixed point. 

The 4d central charges associated to this solution are readily computed from \eqref{achol}, giving:
\equ{\label{acD3ads7}
\boxed{
a_{4d}=c_{4d}=\frac{\pi \, e^{2g_{0}+3f_{0}}\, \vol(\Sigma_{\fg})}{8G_{N}^{(7)}}=\frac{63}{512}\,(\fg-1)\, a_{6d}
}
}
This result is in perfect agreement with the large $N$ limit of the field theory calculation in \eqref{univ rel 64 large N}. This is a nice consistency check that the supergravity solution at hand is dual to the universal RG flow across dimensions discussed around  \eqref{univ rel 64 large N}.

The entire solution corresponds to a BPS 3-brane asymptotic to AdS$_{7}$, whose entropy density is given by \eqref{entp D3} and can be expressed in terms of UV field theory data by using \eqref{acD3ads7}.

\paragraph{AdS$_3$ vacua.}

The Ansatz in this case is
\eqss{
ds^{2}&=e^{2f(r)}(-dt^{2}+dz^{2}+dr^{2})+e^{2g(r)}\,ds^{2}_{M_{4}}\,,\\
F&=-\frac18 \, \omega_{M_{4}}\,,
}
where $ds^{2}_{M_{4}}$ is a constant-curvature metric on a K\"ahler-Einstein manifold $M_{4}$, with K\"ahler form  $\omega_{M_{4}}$, normalized such that 
\equ{
R_{\mu\nu}^{(4)}=-\,g_{\mu\nu}^{(4)}\,,
}
where we took $M_{4}$ to be negatively curved (only this case leads to regular supergravity solutions). A solution of this form within 7d maximally supersymmetric supergravity was found in \cite{Gauntlett:2000ng} (see Equation  (4.12) there and recall  we set $m=2$). As argued above, for this Ansatz the same BPS equations hold in minimal supergravity, which read:\footnote{These can also be obtained as a special case of the general flows studied in  \cite{Benini:2013cda} (see Equation (6.16)  in that reference and set $\lambda_{1}=\lambda_{2}\equiv \lambda$ and $z=0$).}   
\eqss{
e^{-f}f'&=-\frac15\(4e^{-2\lambda}+e^{8\lambda}\)-\frac{3}{160}\,e^{-4\lambda-4g}-\frac{1}{10}\,e^{-2g+2\lambda}\,,\\
e^{-f}g'&=-\frac15\(4e^{-2\lambda}+e^{8\lambda}\)+\frac{1}{80}\,e^{-4\lambda-4g}+\frac{3}{20}\,e^{-2g+2\lambda}\,,\\
e^{-f}\lambda'&=-\frac25\(e^{-2\lambda}-e^{8\lambda}\)+\frac{1}{160}\,e^{-4\lambda-4g}-\frac{1}{20}\,e^{-2g+2\lambda}\,.
}
At the AdS$_{3}$ fixed point the metric functions take the form \eqref{IRfp}, with  
\equ{
e^{5f_{0}}=\frac{1}{24}\,,\qquad e^{10g_{0}}=\frac{27}{2^{16}}\,,\qquad e^{10\lambda_{0}}=\frac43\,,
}
where $\lambda_{0}$ is the value of the scalar field at the fixed point. 

The corresponding 2d central charges can be computed from \eqref{achol} and read:
\equ{\label{crcld1Ads7}
\boxed{
c_{r}=c_{l}=\frac{3e^{f_{0}+4g_{0}}\,\vol(M_{4})}{2G_{N}^{(7)}}=\frac{21\vol(M_{4})}{256\pi^{2}}\,a_{6d}
}}
This result is in nice agreement with the large $N$ limit of the field theory calculation in \eqref{crcl64largeN}.

The entire solution describes a BPS 1-brane asymptotic to AdS$_{7}$, whose entropy density can be expressed in terms of UV SCFT data by combining   \eqref{entp D1} and \eqref{crcld1Ads7}.

\paragraph{AdS$_4$ vacua.}

The relevant AdS$_4$ solution in 7d minimal supergravity was found in \cite{Pernici:1984nw} (see also the later work \cite{Acharya:2000mu}) and preserves only two Poincar\'e and two superconformal supercharges, i.e. the dual 3d SCFT has $\mathcal{N}=1$ supersymmetry. The universal nature of this solution was also emphasized in \cite{Apruzzi:2015wna}. We do not discuss this further here.

\subsubsection{$\cN=4$}
\label{sec:maximal 7d supergravity}

The maximally supersymmetric $SO(5)$ gauged supergravity of \cite{Pernici:1984xx} is the relevant setting for studying twisted compactifications of  6d $\cN=(2,0)$ SCFTs holographically.For topological twists involving an Abelian subgroup of $SO(5)$ it is sufficient to restrict to the $U(1)^{2}$ truncation of  the $SO(5)$ gauged supergravity with bosonic fields the metric, two Abelian gauge fields $A^{(1)},A^{(2)}$ in the Cartan of  $SO(5)$, and two real scalars $\lambda_{1},\lambda_{2}$.  This  truncation was studied in \cite{Liu:1999ai} (see Equations (2.5), (2.12), and (2.13) there for the bosonic Lagrangian and supersymmetry variations).

One may embed the 7d $\cN=2$ gauged supergravity into the $\cN=4$  theory, and the solutions discussed in the previous section are also solutions of the maximally supersymmetric theory, with the additional fields set to zero. From the field theory perspective this corresponds to applying the universal twist of general 6d $\cN=(1,0)$ theories to the special case of $\cN=(2,0)$ theories. In this section, we discuss additional twists possible for $\cN=(2,0)$ theories and their gravity duals. 

\paragraph{AdS$_{5}$ vacua}

The solution describing the twisted compactification of the 6d $\cN=(2,0)$ theory to a 4d $\cN=2$ theory was first found by Maldacena and N\'u\~nez in \cite{Maldacena:2000mw}.  For completeness, we reproduce the answer here, following the notation and conventions in \cite{Bah:2012dg}. The Ansatz is 
\eqss{
ds^{2}&=e^{2f(r)}\,(-dt^{2}+dr^{2}+dz_{1}^{2}+dz_{2}^{2}+dz_{3}^{2})+e^{2g(r)}\,ds^{2}_{\Sigma_{\fg>1}}\,,\\
F^{(1)}&=-\frac14\, \dvol_{\Sigma_{1}}\,,\qquad F^{(2)}=0\,.
}
At the AdS$_{5}$ fixed point the metric functions take the form \eqref{IRfp}, with (see Equation  (3.8) in \cite{Bah:2012dg}):
\equ{
e^{2f_{0}}=2\,e^{2g_{0}}=\frac{1}{2^{4/5}}\,,\qquad e^{2\lambda_{1}^{(0)}}=\frac12 \,e^{2\lambda_{2}^{(0)}}= \frac{1}{2^{3/5}}\,,
}
where $\lambda_{1,2}^{(0)}$ are the values of the scalars at the fixed point.  

Using \eqref{achol},  the corresponding 4d central charges read  
\equ{\label{a4da6dmax}
\boxed{
a_{4d}=c_{4d}= \frac{\pi e^{3f_{0}+2g_{0}}\,\vol({\Sigma_{\fg}}) }{8G_{N}^{(7)}} =\frac{7}{48}\,(\fg-1)\,a_{6d}
}
}
which exactly reproduces the large $N$ field theory result \eqref{univ rel 64 large N z=1}. 

The entire solution describes a BPS 3-brane in  AdS$_{7}$ and its entropy density is readily expressed in terms of UV data by combining \eqref{entp D3}  and \eqref{a4da6dmax}.

\paragraph{AdS$_{3}$ vacua.}

The compactification on $M_{4}=\Sigma_{1}\times \Sigma_{2}$, with both Riemann surfaces of negative curvature,  preserving 2d $\cN=(2,2)$ supersymmetry is given by (see Equation  (5.26) and Appendix G of \cite{Benini:2013cda})
\eqss{
ds^{2}&=e^{2f(r)}\,(-dt^{2}+dr^{2}+dz^{2})+e^{2g_{1}(r)}\,ds^{2}_{\Sigma_{1}}+e^{2g_{2}(r)}\,ds^{2}_{\Sigma_{2}}\,,\\
F^{A}&=-\frac14\,\dvol_{\Sigma_{1}}\,,\qquad F^{B}=-\frac14\,\dvol_{\Sigma_{2}}\,.
}
At the AdS$_{3}$ fixed point the metric functions take the form \eqref{IRfp}, with   
\equ{
e^{2f_{0}}=e^{2g_{1}^{(0)}}=e^{2g_{2}^{(0)}}=\frac14\,,
}
and the scalars vanish: $\lambda_{1}=\lambda_{2}=0$.

The corresponding 2d central charges, computed from \eqref{achol}, read:
\equ{\label{c2da6dmax}
\boxed{
c_{r}=c_{l}=\frac{3e^{f_{0}+2g_{1}^{(0)}+2g_{2}^{(0)}}}{2G_{N}^{(7)}}\,\vol(\Sigma_{1})\,\vol(\Sigma_{2})=\frac74\,(\fg_{1}-1)(\fg_{2}-1)\,a_{6d}
}
}
which exactly reproduces the field theory result \eqref{c2dn=2022}.

The entire solution describes a BPS 1-brane asymptotic to AdS$_{7}$, whose entropy density is given in terms of UV data by combining  \eqref{entp D1} and \eqref{c2da6dmax}.

\paragraph{AdS$_4$ vacua.}

AdS$_4$ vacua arising from M5-branes wrapping special Lagrangian 3-cycles were found in Section~3.1 of \cite{Gauntlett:2000ng} (see also \cite{Acharya:2000mu}). These describe the twisted compactification of 6d $\cN=(2,0)$ theories on three-manifolds, discussed at the end of Section~\ref{subsec:6dN20CFT}. For $M_{3}$ an Einstein space of negative constant curvature, there is an  AdS$_{4}$ fixed point, for which the metric functions take the form \eqref{IRfp}, with
\equ{
e^{f_{0}}=\frac{e^{4\lambda_{0}}}{2}\,, \qquad e^{2g_{0}}=\frac{e^{8\lambda_{0}}}{8}\, \qquad e^{10\lambda_{0}}=2\,,
}
where $\lambda_{0}$ is the value of the only non-trivial scalar field at the fixed point.

The  $S^{3}$ free energy of the corresponding  AdS$_{4}$ solution follows from  \eqref{Fgrav S3}  and is given by
\equ{\label{D2AdS7}
\boxed{
F_{S^{3}}=\frac{\pi e^{2f_{0}+3g_{0}}\,\vol(M_{3})}{2G_{N}^{(7)}}=\frac{7\,\vol(M_{3})}{96\pi\sqrt {2}}\,a_{6d}
}
}
where in the last equality we used \eqref{achol}. The entire solution describes a BPS 2-brane asymptotic to AdS$_{7}$, whose entropy density is given in terms of UV data by combining  \eqref{entp D2} and \eqref{D2AdS7}.

\paragraph{AdS$_2$ vacua.}

AdS$_2$ vacua arising from M5-branes wrapping special Lagrangian 5-cycles $M_{5}$ were found in Section 3.3 of \cite{Gauntlett:2000ng}. These describe the twisted compactification of 6d $\cN=(2,0)$ theories on five-manifolds discussed at the end of Section~\ref{subsec:6dN20CFT}. Regular solutions to the supergravity BPS equations were found for $M_{5}$ being $S^{5}$ or $\Bbb H^{5}$ with an Einstein metric of normalized curvature $\kappa=1$ and $\kappa=-1$ respectively.\footnote{It is curious to note that this is the only example of a regular universal domain wall solution with a positively curved metric on the compactification manifold $M_{d-p}$.} The supergravity scalars vanish and the metric functions are given by
\begin{equation}
e^{2f_0}= \ds\frac{(2-\kappa)^2}{64}\;, \qquad e^{2g_0}= \ds\frac{(2-\kappa)}{16}\;,
\end{equation}
With this at hand the black hole entropy corresponding to this near horizon AdS$_{2}$ solution can be written as
\equ{
\boxed{
S_{\text{BH}}=\frac{e^{5g_{0}}\, \vol(M_{5})}{4G_{N}^{(7)}}=\frac{7(2-\kappa)^{5/2}\, \vol(M_{5})}{3\times 2^{12}\pi^{2}}\, a_{6d}
}
}
where for the last identity we made use of \eqref{achol}. It would be nice to reproduce this result by a field theory calculation.

\subsection{6d  supergravity}
\label{sec:6d  supergravity}

There is a unique six-dimensional gauged supergravity theory with a supersymmetric AdS$_6$ vacuum that can be constructed out of the gravity multiplet. This was done by Romans in  \cite{Romans:1985tw}. The theory has 16 supercharges and the AdS$_6$ vacuum is invariant under the supergroup $F(4)$, which is also the unique superconformal group in 5d.  The bosonic field content of the theory is given by the graviton $g_{\mu\nu}$, an $SU(2)$ gauge potentials $A_{\mu}^{I}$, an Abelian one-form potential $A_{
\mu}$, a two-index tensor gauge field $B_{\mu\nu}$, and a scalar $\phi$. The two-form ``eats'' the one-form $A_{\mu}$ and becomes massive, which can be implemented by choosing $A_{\mu}=0$. This bosonic content mimics precisely the bosonic operators in the energy-momentum multiplet of 5d SCFTs. In particular, the $SU(2)$ gauge field $A_{\mu}^{I}$ is dual to the R-current. The fermionic field content is four gravitinos $\psi_{\mu\, i}$ and four gauginos $\chi_{i}$ with a symplectic Majorana condition. To implement the topological twist of interest here we only need to use the metric, the gauge field, and the scalar. Therefore we will set the two-form $B_{\mu\nu}=0$ from now on. Using the conventions summarized in \cite{Naka:2002jz}, the bosonic Lagrangian reads\footnote{Compared to \cite{Naka:2002jz} we have defined $\varphi_{\text{here}}=\frac{1}{\sqrt 2}\, \phi_{\text{there}}$ and denote the coupling constant by $\bar g$ instead of $g$. } 
\eqss{
e^{-1}\mathcal L&=R-4\,(\partial_{\mu}\varphi)^{2}-e^{-2 \varphi}\,F^{I\, \mu\nu}
F^{I}_{\mu\nu}-V(\varphi)\,,\\
V(\varphi)&=\frac12 \(m^{2}\,e^{-6 \varphi}-\bar g^{2}\,e^{2 \varphi}-4\bar g m \, e^{-2 \varphi}\)\,,
}
where $\bar g$ is the $SU(2)$ coupling constant and $m$ is a mass parameter associated to  $B_{\mu\nu}$. Depending on the signs of $\bar g,m\,,$ or if any of these parameters vanishes, this Lagrangian actually describes five different theories. Here we are interested in the case $\bar g>0,m>0$, which corresponds to the theory labelled  $\cN=4^{+}$ in \cite{Romans:1985tw}. Further setting $\bar g=3m$ the theory admits an AdS$_{6}$ vacuum with $F(4)$ symmetry, corresponding to the extremum of the potential at $\varphi=0$ where it takes the value $V|_{\varphi=0}=-10\,m^{2}$.  Comparing this to \eqref{normalization LAdS} we set $m=\sqrt 2$ in what follows to normalize  $L_{\text{AdS}_{6}}=1$.

Our main interest here is in the solutions of this $F(4)$ gauged supergravity describing the twisted compactification of 5d SCFTs on two-, three-, and four-manifolds, as briefly discussed at the end of Section~\ref{sec:Comments on 3d and 5d SCFTs}. The corresponding AdS$_4$, AdS$_{3}$, and AdS$_{2}$  vacua, which we review  below,  were constructed in \cite{Nunez:2001pt,Naka:2002jz}. We mostly follow the conventions in \cite{Naka:2002jz}, where the relevant BPS equations are written (see Equations (5.12) there). We denote by $D=5-p$ the dimension of the compactification manifold. The supergravity equations imply the metric on this manifold should be  Einstein,  whose normalized curvature we denote by $\kappa$. With this notation the BPS equations read
\eqss{\label{BPSNaka}
e^{-f}\varphi'&=\frac{1}{4}\(3e^{\varphi}-3 e^{-3\varphi}-\frac{\kappa \,D}{6}\,e^{-2g-\varphi}\)\,,\\
e^{-f}g'&=-\frac{1}{4}\(3e^{\varphi}+ e^{-3\varphi}+\frac{\kappa \, (8-D)}{6} \,e^{-2g-\varphi}\)\,,\\
e^{-f}f'&=-\frac{1}{4}\(3e^{\varphi}+e^{-3\varphi}-\frac{\kappa \,D}{6} \,e^{-2g-\varphi}\)\,.
}
Next, we describe the asymptotic behavior of solutions to these equations for the relevant values of $D$. 

\paragraph{AdS$_4$ vacua.}

The twisted compactification of a 5d SCFT on a Riemann surface to a 3d SCFT is described by a solution of the form
\eqss{
ds^{2}&=e^{2f(r)}(-dt^{2}+dz_{1}^{2}+dz^{2}_{2}+dr^{2})+e^{2g(r)}ds^{2}_{\Sigma_{\fg}}\,,\\
F^{I}&=\frac{2\kappa}{3\sqrt 2}\,\delta^{I3} \, \dvol_{\Sigma_{\fg}}\,.
}
Setting $D=2$ in \eqref{BPSNaka} one finds that for $\kappa=-1$ there is an AdS$_{4}$ fixed point, with metric functions of the form \eqref{IRfp}, with 
\eqss{
e^{f_{0}}=\frac{2}{3}\,e^{-\varphi_{0}}\,,\qquad  e^{g_{0}}=\frac{\sqrt2}{3}\,e^{-\varphi_{0}}\,,\qquad e^{-4\varphi_{0}}=\frac32\,,
}
with $\varphi_{0}$ the value of the scalar field at the AdS$_4$ fixed point.

The free energy on $S^{3}$ of the 3d $\cN=2$ SCFT dual to this AdS$_4$ vacuum is computed from \eqref{Fgrav S3} by evaluating the on-shell action and reads
\equ{
\boxed{
\label{FS3FS5}
F_{S^{3}}=\frac{\pi e^{2f_{0}+2g_{0}}\,\vol(\Sigma_{\fg})}{2 G_{N}^{(6)}}=-\frac{8}{9}\,(\fg-1)\, F_{S^{5}}
}
}
where we have also used \eqref{Fgrav S3}.  This universal relation is a nontrivial prediction of supergravity which would be interesting to derive directly in field theory. 

The entire solution describes a  2-brane asymptotic to AdS$_{6}$, whose entropy density is given in terms of UV data by combining \eqref{entp D2}  and \eqref{FS3FS5}.

\paragraph{AdS$_3$ vacua.}

The solution describing the twisted compactification on a hyperbolic three-manifold $M_{3}=\Bbb H^{3}/\Gamma$ was constructed in Section 3.1 of \cite{Nunez:2001pt} and is of the form 
\eqss{
ds^{2}&=e^{2f(r)}(-dt^{2}+dz^{2}+dr^{2})+e^{2g(r)}ds^{2}_{\Bbb H^{3}}\,,\\
A^{1}&=-\frac{1}{3\sqrt 2}\frac{dx_{1}}{x_{2}}\,, \qquad A^{3}=-\frac{1}{3\sqrt 2}\frac{dx_{3}}{x_{2}}\,, \qquad A^{2}=0\,.
}
where $ds^{2}_{\Bbb H^{3}}=\frac{1}{x_{2}^{2}}(dx_{1}^{2}+dx_{2}^{2}+dx_{3}^{2})$ is the metric on hyperbolic space, which we quotient by an appropriate discrete group. Setting $D=3$ and $\kappa=-1$ in \eqref{BPSNaka} one finds an AdS$_{3}$ fixed point, where the metric functions take the form \eqref{IRfp}, with   
\equ{
e^{f_{0}}=\frac{1}{2}\,e^{-\varphi_0}\,, \qquad e^{g_{0}}=\frac{1}{\sqrt 6}\,e^{-\varphi_0} \,, \qquad e^{-4\varphi_{0}}=2\,,
}
with $\varphi_0$  the value of the scalar at the horizon.

This twisted compactification preserves 2d $\cN=(1,1)$ supersymmetry and thus we have fewer technical tools to study the IR SCFT. However, we can compute holographically the central charge of the 2d theory using \eqref{achol} and find 
\equ{
\boxed{\label{crclFS3}
c_{r}=c_{l}=\frac{3e^{f_{0}+3g_{0}}\, \vol(M_{3})}{2G_{N}^{(6)}}=-\sqrt{\frac32}\,\frac{\vol(M_{3})}{4\pi^{2}}\,F_{S^5}
}
}
where in the last equality we have used \eqref{Fgrav S3}.

The entire solution describes a 1-brane asymptotic to AdS$_{6}$, whose entropy density is given in terms of UV data by using from the formulas \eqref{entp D1} and \eqref{crclFS3}.

\paragraph{AdS$_2$ vacua.}

AdS$_2$ vacua were found in \cite{Naka:2002jz}, corresponding to twisted compactification of 5d SCFTs on four-manifolds $M_4$.   Since the gauge field in the supergravity theory is only $SU(2)$, this restricts the possible 4-manifolds one may consider. One option is for $M_{4}$ to be K\"ahler.\footnote{Another option is for $M_{4}$ to be a co-associative cycle in a non-compact $G_{2}$ holonomy manifold, in which case the metric has the same form; see Section~5.1 of \cite{Naka:2002jz}. } Then, setting $D=4$ in  \eqref{BPSNaka}  the BPS equations dictate that $\kappa=-1$ and
one  finds an AdS$_{2}$ fixed point, where the metric functions take the form \eqref{IRfp}, with  
\equ{
e^{f_{0}}=\frac13\,e^{-\varphi_0}\,, \qquad e^{g_{0}}=\frac13\,e^{-\varphi_0} \,, \qquad e^{-4\varphi_{0}}=3\,,
}
with $\varphi_{0}$ the value of the scalar at the horizon. 

The entire spacetime is a BPS black hole asymptotic to AdS$_6$, with entropy
\equ{\label{SBHFS5}
\boxed{
S_{\text{BH}}=\frac{e^{4g_0} \, \vol(M_4)}{4G_N^{(6)}}=- \frac{\vol(M_4)}{36\pi^{2}}\,F_{S^5}
}
}
where in the last equality we used \eqref{Fgrav S3}. This is  another universal prediction from holography which would be interesting to test with field theory methods, by comparing the partition function on $S^{1}\times M_{4}$ (with a universal topological twist on $M_{4}$) and the partition function on $S^{5}$ at large $N$.

\subsection{5d supergravity}
\label{sec:5d SUGRA}

We now proceed to study AdS$_3$ and AdS$_2$ vacua of 5d $\mathcal{N}=2$ and  $\mathcal{N}=4$ minimal gauged supergravity. These describe universal twisted compactifications of  4d $\cN=1$ and $\cN=2$ SCFTs on Riemann surfaces and three manifolds. 

\subsubsection{$\cN=2$}
\label{sec:N=2 supergravity}

This theory has 8 supercharges and bosonic field content the metric $g_{\mu\nu}$ and a $U(1)$ gauge field $A_{\mu}$ dual to the stress-energy tensor and the R-current in the dual 4d $\cN=1$ SCFT.\footnote{This theory can be obtained from the well-studied STU model of gauged supergravity by setting the vector multiplet fields to zero} Here we follow the conventions of \cite{Maldacena:2000mw,Benini:2013cda} in which the Lagrangian reads (see Equation  (46) in \cite{Maldacena:2000mw})
\equ{
e^{-1}\mathcal L=R+12-\frac{3}{4}\,F_{\mu\nu}^{2}+\frac{1}{4}\,e^{-1}\epsilon^{\mu\nu\alpha\beta\rho}F_{\mu\nu}F_{\alpha\beta}A_{\rho}\,,
}
where  the cosmological constant is normalized such that $L_{\text{AdS}_{5}}=1$.

\paragraph{AdS$_3$ vacua.}

We are interested in BPS solutions of the supergravity theory with AdS$_{5}$ asymptotics and near-horizon geometry AdS$_{3}\times \Sigma_{\fg}$. These describe the universal flow from 4d $\cN=1$ SCFTs to 2d $\cN=(0,2)$ discussed in  SCFTs \cite{Benini:2015bwz,Bobev:2014jva} and reviewed in Section~\ref{N=1 Sigma}.

The Ansatz for the metric and gauge field are
\eqss{
ds^{2}&=e^{2f(r)}\, (-dt^{2}+dz^{2}+dr^{2})+e^{2g(r)}\, ds^{2}_{\Sigma_{\fg}}\,,\\
F&=\frac\kappa3 \, \dvol_{\Sigma_{\fg}}\,,
}
and the resulting BPS equations read:\footnote{As noted above $\cN=2$ gauged supergravity can be obtained by setting $A_{\mu}^{1}=A_{\mu}^{2}=A_{\mu}^{3}\equiv A_{\mu}$ and $\phi_{1}=\phi_{2}=0$ in the $U(1)^{3}$ truncation of $\cN=8$ gauged supergravity. Thus it is useful to make contact with previous references on this well-studied model. Setting $a_{1}=a_{2}=a_{3}=\frac13$ and $\phi_{1}=\phi_{2}=0$ in Equation  (3.20) in \cite{Benini:2013cda} leads to the BPS equations in \eqref{eq:BPSads35dN2}. These also coincide with equations (65-68) in \cite{Maldacena:2000mw} after setting $a=b=c=\frac13$ and $\varphi=0$.}
\eqss{\label{eq:BPSads35dN2}
e^{-f}f'&=-1+\frac\kappa6\,e^{-2g} \,\\
e^{-f}g'&=-1-\frac\kappa3\,e^{-2g} \,.
}
The solution to these equations with the required asymptotics, which exists only for $\kappa=-1$, describes a magnetically charged BPS black string in AdS$_{5}$ and was discussed in \cite{Klemm:2000nj}. The entire domain wall solution preserves 2 real supercharges, which is enhanced as usual  to 4 supercharges at the horizon. 

For the purpose of computing the central charge of the IR 2d SCFT, we focus on the AdS$_{3}$ fixed point of  \eqref{eq:BPSads35dN2}, where the metric functions take the form \eqref{IRfp}, with
\equ{
e^{f_{0}}=2\,,\qquad e^{2g_{0}}=\frac23\,.
}
The 2d central charges can then be easily be found to read 
\equ{\label{crcla4d}
\boxed{
c_{r}=c_{l}=\frac{3e^{f_{0}+2g_{0}}\, \vol(\Sigma_{\fg})}{2G_{N}^{(5)}}=\frac{32}{3}\,(\fg-1)\,a_{4d}
}
}
reproducing the universal field theory result in \eqref{4d02univ}.

The entropy density of the black string can be written in terms of the data of the 4d UV SCFT by combining    \eqref{entp D1} with \eqref{crcla4d}.

\paragraph{AdS$_2$ vacua.}

Since the minimal gauged supergravity in 5d contains only a $U(1)$ gauge field and the holonomy group of a generic Riemannian three-manifold is $SO(3)$, we do not expect to find supersymmetric AdS$_2$ vacua. However, it is possible to construct such vacua in the 5d $\cN=4$ theory, which we review next. 

\subsubsection{$\cN=4$}
\label{sec:5d N=4 supergravity}

The 5d $\cN=4$ gauged supergravity theory has 16 supercharges. The bosonic field content is the graviton, $g_{\mu\nu}$, $SU(2)\times U(1)$ gauge fields $A_{\mu}^{I}$, $I=1,2,3$, and $A_{\mu}$,  two antisymmetric tensor fields $B_{\mu\nu}^{1,2}$, and a scalar $\phi$. This structure nicely corresponds to the one of the stress-energy tensor multiplet in the dual 4d $\cN=2$ SCFT. Depending on the values of the  $U(1)$ and $SU(2)$ gauge  coupling constants $g_1,g_2$, respectively, there are three different gauged supergravity theories found in \cite{Romans:1985ps}.  Here we are interested in the unique theory, denoted by $\cN=4^+$ in \cite{Romans:1985ps}, for which there is a supersymmetric AdS$_5$ critical point. To obtain that one has to set $\bar g\equiv g_2=\sqrt2 g_1$ in the notation of \cite{Romans:1985ps}. Here we are interested in supersymmetric solutions of this theory which describe the topological twisted compactifications discussed in  Section~\ref{sec:N=2 on Sigma} and thus we set the two-form fields $B_{\mu\nu}^{1,2}$ to zero to find the following bosonic Lagrangian\footnote{We obtain this by setting all fermionic fields and antisymmetric tensor fields to zero in the Lagrangian (2.14) in  \cite{Romans:1985ps} and sending $g_{\mu\nu}\to -g_{\mu\nu}$ to change to a ``mostly plus'' signature. We have also rescaled the scalar field $\phi_{\text{there}}=\frac12 \phi_{\text{here}}$ and the $SU(2)$ gauge field $F^{I}_{\text{here}}=\sqrt{2} F^{I}_{\text{there}}$. }
\equ{\label{Lagrangian 5d N=4+}
e^{-1}\mathcal L=R-\frac12\,(D_\mu \phi)^2-e^{-\frac{4\phi}{\sqrt6}}\,F^{\mu\nu}F_{\mu\nu}-\frac{1}{2}\,e^{\frac{2\phi}{\sqrt6}}\,F^{\mu\nu\, I}F_{\mu\nu}^{I}+\frac{1}{2}\, \bar g^2\, (2e^{\frac{\phi}{\sqrt6}}+e^{-\frac{2\phi}{\sqrt6}})\,.
}
Normalizing the scale of AdS$_5$ as  $L_{\text{AdS}_{5}}=1$ requires to set the value of the potential at the critical point $\phi=0$ to be $V|_{\phi=0}=-\frac{3\bar g^{2}}{2}=-12$ and hence $\bar g=2\sqrt 2$. We use this normalization from now on and for convenience define $\phi \equiv \sqrt{6}\varphi$.

\paragraph{AdS$_3$ vacua.}

We are after a solution that describes the universal flow from a 4d $\cN=2$ theory to a 2d $\cN=(2,2)$ theory. As discussed in Section~\ref{sec:N=2 on Sigma} this is obtained by a topological twist along the Cartan of $SU(2)$. Thus, we are after AdS$_{3}$ solutions where only the gauge field along the Cartan of $SU(2)$ is turned on and the $U(1)$ gauge field vanishes.
The Ansatz for the metric and gauge fields is thus
\eqss{
ds^{2}&=e^{2f(r)}(-dt^{2}+dz^{2}+dr^{2})+e^{2g(r)}ds^{2}_{\Sigma_{\fg}}\,,\\
F^{I}&=\frac{\kappa}{2}\, \delta^{I3} \, \dvol_{\Sigma_{\fg}}\,, \qquad F=0\,.
}
The corresponding BPS equations read
\eqss{\label{BPS eqs 4dN=2 2dN=22}
e^{-f}f'&=-\frac13 \, (2e^{-\varphi}-e^{2\varphi})+\frac\kappa6 \, e^{-2g+\varphi}\,,\\ 
e^{-f}g'&=-\frac13\, (2e^{-\varphi}-e^{2\varphi})-\frac\kappa3 \,e^{-2g+\varphi}\,,\\
e^{-f}\varphi'&=-\frac23\,(e^{-\varphi}+e^{2\varphi})+\frac\kappa6 \, e^{-2g+\varphi}\,.
} 
The solution to these equations with the desired asymptotics, which exists only for $\kappa=-1$, was  constructed in \cite{Romans:1985ps}.\footnote{Indeed, setting $x=\frac12$ in \cite{Romans:1985ps} corresponds to turning off the $U(1)$ gauge field and leaving only a non-zero gauge field for the Cartan of $SU(2)_{R}$, see Equation (4.5) in \cite{Romans:1985ps}.}  At the AdS$_{3}$ fixed point, the metric functions take the form \eqref{IRfp}, with  
\equ{
 e^{2f_{0}}=e^{2g_{0}}=\frac{1}{2^{4/3}}\,,\qquad e^{\varphi}=2^{1/3}\,.
}
The entire domain wall solution preserves 4 real supercharges which are enhanced to 8 supercharges at the horizon. The corresponding 2d central charge is given by
\equ{\label{crcla4dmax}
\boxed{
c_{r}=c_{l}=\frac{3L_{\text{AdS}_{3}}}{2G_{N}^{(3)}}=\frac{3e^{f_{0}+2g_{0}}\,\vol(\Sigma_{\fg})}{2G_{N}^{(5)}}=12\,(\fg-1)\, a_{4d}\,
}
}
which nicely matches the large $N$ field theory result \eqref{4d2d2,2largeN}. The entropy density of this supersymmetric black string solution in terms of UV SCFT data is obtained by combining  \eqref{crcla4dmax} and \eqref{entp D1}.

We note that the BPS equations \eqref{BPS eqs 4dN=2 2dN=22} coincide with the BPS equations describing the twisted compactification of $\cN=4$ SYM on a Riemann surface preserving 2d $\cN=(2,2)$ supersymmetry discussed in \cite{Maldacena:2000mw}  (see Equations (14)-(16) there).\footnote{These can also be obtained by setting $a_{1}=a_{2}=-\frac\kappa2$ and $a_{3}=0$ in Equation  (3.20) of \cite{Benini:2013cda}.}  In that reference,  the flow was described within the $U(1)^3$ truncation of $\cN=8$ supergravity, which consists of the metric $g_{\mu\nu}$, three Abelian gauge fields $\cA_\mu^{1,2,3}$ in the Cartan of $SO(6)$ and two neutral scalars $\phi_1$ and $\phi_2$. The solution described above corresponds to a solution with $\cA_\mu^1=\cA_\mu^2$, $\cA_\mu^3=\phi_2=0$, and identifying the remaining scalar $\phi_1=\sqrt 6\varphi$ in which case one can see that the Lagrangians and supersymmetry transformations coincide.\footnote{One can check that setting $\cA_\mu^1=\cA_\mu^2$, $\cA_\mu^3=\phi_2=0$, and  $\phi_1=\sqrt 6\varphi$ in the Lagrangian given in Equation  (46) in \cite{Maldacena:2000mw}, it matches \eqref{Lagrangian 5d N=4+}, assuming only the  Cartan of $SU(2)$ is excited and setting $\bar g=2\sqrt2$.} These results are in harmony with the fact that the $\cN=4^{+}$ Romans supergravity theory can be obtained also as a truncation of the 5d $SO(6)$ maximally supersymmetric gauged supergravity of \cite{Gunaydin:1985cu}.

The twisted compactification corresponding to the $\beta$-twist discussed in Section \ref{sec:N=2 on Sigma} is realized in the $\cN=4^{+}$ supergravity by turning on magnetic flux for the $U(1)$ gauge field, $A_{\mu}$, and switching off the $SU(2)$ gauge field flux, $A_{\mu}^{I}$. However one can show that an Ansatz with this field configuration leads to a singular supergravity flow solution which does not flow to an AdS$_3$ vacuum in the IR. This suggests that the dual 2d $(0,4)$ theory in the IR has some pathology, for example an accidental symmetry or a non-normalizable vacuum state.

\paragraph{AdS$_2$ vacua.}

AdS$_2$ vacua in 5d $\mathcal{N}=4$ minimal gauged supergravity were found \cite{Nieder:2000kc}. This describes the twisted compactification of a 4d $\cN=2$ theory on $M_3=\Bbb H^3$. Since the structure group of the three-manifold is $SO(3)$ to implement the topological twist one has to switch off the Abelian gauge field in the supergravity theory. The Ansatz for the metric and $SU(2)$ gauge field is
\eqss{\label{AdS25dsugra}
ds^2&=e^{2f(r)}\,(-dt^2+dr^2)+e^{2g(r)}\, ds^2_{\Bbb H^3}\\
A^1&=\frac{1}{2\sqrt 2}\, \cosh \phi \, d\theta\,, \quad A^2=\frac{1}{2\sqrt 2}\, \cos \theta \, d\nu\,,\quad A^3=-\frac{1}{2\sqrt 2}\,\sin \theta \cosh \phi \, d\nu\,,
}
where $ds^2_{\Bbb H^3}=d\phi^2+\sinh^2 \phi\,(d\theta^2+\sin^2\theta \, d\nu^2)$. The corresponding BPS equations were derived in  \cite{Nieder:2000kc} (see Equation  (56) there), which we reproduce here for completeness:
\eqss{
e^{-f}f'&= -\frac12 \,e^{\varphi-2g}-\frac{1}{3}\,(2e^{-\varphi}+e^{2\varphi})\,,\\
e^{-f}g'&= \frac12\,e^{\varphi-2g}-\frac{1}{3}\,(2e^{-\varphi}+e^{2\varphi})\,,\\
e^{-f}\varphi'&=-\frac12 \, e^{\varphi-2g}-\frac{2}{3}\,(e^{-\varphi}-e^{2\varphi})\,.
}
As discussed in \cite{Nieder:2000kc} (see Equation  (42) there), at the AdS$_2$ fixed point  the metric functions take the form \eqref{IRfp}, with 
\equ{
e^{2g_{0}}=\frac{1}{4^{1/3}}\,,\qquad e^{2f_{0}}=\frac{1}{4^{4/3}}\,, \qquad e^{3\varphi_{0}} =4\,, 
}
with $\varphi_{0}$ the value of the scalar at the horizon.  The domain wall solution which interpolates between AdS$_5$ and this AdS$_2$ vacuum preserves two real supercharges (enhanced to 4 supercharges in the near horizon limit) and can be thought of as a BPS black hole with a hyperbolic horizon. The entropy of the black hole is given by
\equ{\label{S1H3BH}
\boxed{
S_{\text{BH}}=\frac{e^{3g_0}\, \vol(\Bbb H^3)}{4G_N^{(5)}}=\frac{\vol(\Bbb H^3)}{\pi}\,a_{4d}
}
}
It would be interesting to establish this universal relation using field theory methods.

\subsection{4d supergravity}
\label{sec:4d supergravity}

Here we discuss black hole solutions in four-dimensional gauged supergravity describing the universal twisted compactification of 3d $\cN=2$ and $\cN=4$ SCFTs on a Riemann surface. The field theory setting was briefly discussed in Section~\ref{sec:Comments on 3d and 5d SCFTs} and  we refer to \cite{ABCMZ} for more details. The study of asymptotically AdS$_{4}$ back holes has received renewed attention, following the discovery of the 3d superconformal theories describing the worldvolume of M2-branes and their  AdS$_{4}$ duals \cite{Aharony:2008ug}; see  \cite{Cacciatori:2009iz,DallAgata:2010ejj,Hristov:2010ri} and references thereof. The interpretation of some of these solutions as twisted compactifications of the ABJM theory was provided in \cite{Benini:2015eyy,Benini:2016rke} (see also \cite{Gauntlett:2001qs} for earlier work), where the microsocpic entropy of these black holes was reproduced using the supersymmetric index defined in \cite{Benini:2015noa}. In this section, we focus on universal solutions describing the twisted compactification of a large class of 3d $\cN=2$ theories. This was recently used in \cite{ABCMZ} as a tool to count the black hole microstates for a large class of theories with M-theory as well as massive IIA duals  (see also \cite{Benini:2017oxt,Hosseini:2017fjo} for non-universal examples in massive IIA).

\subsubsection{$\cN = 2$ }
\label{subsubsec:4dN2SG}

We are interested in asymptotically AdS$_4$ black hole solutions in minimal $\cN = 2$ gauged supergravity which preserve two supercharges. The near-horizon geometry is AdS$_2 \times \Sigma_{\mathfrak{g}}$ and the entire solution describes the holographic RG flow from a 3d $\cN=2$ SCFT on $\Sigma_{\fg}$ to a 1d superconformal quantum mechanics. Solutions in non-minimal gauged supergravity were summarized in \cite{Benini:2015eyy} where references to the extensive earlier literature on the subject can also be found. The minimal theory is obtained by setting $\mathfrak n_{a}=\frac{\kappa}{2}\,, \vec \phi=0\,, L_{a}=1$ in \cite{Benini:2015eyy}. The Lagrangian reads
\equ{
e^{-1}\mathcal L=R-2 F_{\mu\nu}^{2}-V\,,
}
where $V=-12\bar g^{2}$. Comparing to \eqref{normalization LAdS}  we set $\bar g=1/\sqrt 2$ to normalize $L_{\text{AdS}_{4}}=1$.

The Ansatz of interest takes the form
\eqss{
ds^{2}&=e^{2f(r)}\,(-dt^{2}+dr^{2})+e^{2g(r)}\,ds^{2}_{\Sigma_{\fg}}\,,\\
F&=-\frac{\kappa}{2\sqrt2}\, \dvol_{\Sigma_{\fg}}\,,
}
and the BPS equations read
\eqss{\label{BPSN=2sugra}
e^{-f}f'&=\,-1+\frac{\kappa}{2} \,e^{-2g}\,,\\
e^{-f}g'&=\,-1-\frac{\kappa}{2} \,e^{-2g}\,.
}
For $\kappa=-1$ these equations admit a full analytic solution corresponding to the magnetically charged black hole of \cite{Romans:1991nq,Caldarelli:1998hg}, preserving two supercharges.\footnote{A generalization to include rotation while maintaining supersymmetry was also found in this reference.} The uplift of this solution to eleven dimensions, and its interpretation as wrapped M2-branes on a Calabi-Yau five-fold, was given in \cite{Gauntlett:2001qs}, where  other interesting wrapped membranes solutions were also studied. The uplift to massive IIA is provided in \cite{ABCMZ}. For $\kappa=1$ one finds an IR singularity (see Section 3.4 of  \cite{Gauntlett:2001qs}).

Setting  $\kappa=-1$,  one finds a regular horizon asymptotic to AdS$_2 \times \Sigma_{\mathfrak{g}}$ and  the metric functions take the form \eqref{IRfp}, with 
\equ{\label{solN=2BH}
e^{f_{0}}=e^{2g_{0}}=\frac12\,.
}
The black hole entropy is given by
\equ{\label{BHFS3univ}
\boxed{
S_{\text{BH}}=\frac{e^{2g_{0}}\, \vol(\Sigma_{\fg})}{4G_{N}^{(4)}}=(\fg-1)\,F_{S^{3}}
}
}
where in the last equality we used \eqref{Fgrav S3}. This exactly reproduces  the large $N$ field theory result \eqref{F3d1d}, provided the identification $S_{\text{BH}}=\text{Re} \log Z(\bar \Delta_{I})$, as shown in \cite{ABCMZ} for this class of solutions.\footnote{This was argued to hold more generally for a larger class of black holes with additional charges in \cite{Benini:2015eyy}. }

It is worth highlighting the power behind the rather simple-looking universal relation \eqref{BHFS3univ}. As we have argued on general grounds, this four-dimensional black hole can be uplifted to ten- or eleven-dimensional supergravity in infinitely many ways depending on the choice of six- or seven-manifold; each uplift describes the twisted compactification of a different 3d $\cN=2$ SCFT.  This fact, combined with the universal relation \eqref{Fgrav S3}, was recently used in \cite{ABCMZ} to arrive at a microscopic derivation of the entropy of this infinite family of black holes. Similarly, we expect that the various universal relations derived in this paper will be useful in generalizing these results to even larger classes of black branes in different dimensions.

Non-universal compactifications, i.e., corresponding to turning background flavor fluxes in the SCFT, are described by black holes charged under additional vector multiplets. Such solutions were first considered in \cite{Cacciatori:2009iz} and studied further in a number of papers, notably \cite{DallAgata:2010ejj,Hristov:2010ri} (see \cite{Benini:2015eyy} for a more complete list of references).  The field theory interpretation of these black holes and a microscopic derivation of their entropy was carried out in \cite{Benini:2015eyy,Benini:2016rke} using the twisted index of \cite{Benini:2015noa}. The solution of the minimal theory can be seen as a special case of these, obtained by setting the vector multiplets to zero.

\subsubsection{$\cN = 4$ }

In this section, we investigate which  twisted compactifications of 3d $\cN=4$ SCFTs admit a holographic description. These theories have an $SO(4)\simeq SU(2)_{C}\times SU(2)_{H}$ R-symmetry and the compactification on Riemann surfaces was recently considered in \cite{Gaiotto:2016wcv}. The appropriate supergravity is minimal $\cN = 4$  gauged supergravity  \cite{Das:1977pu}. This supergravity has sixteen supercharges and bosonic content the graviton, an $SO(4)$ gauge field, a dilaton, and an axion. It can be obtained  as an $S^{7}$ reduction of eleven-dimensional supergravity \cite{Cvetic:1999au}. 

Following the notation in \cite{Cvetic:1999au} we  denote the two $SU(2)$ gauge fields by $A$ and  $\tilde A$, the dilaton by $\phi$, and the axion by $\chi$. Setting $\chi=0$  the bosonic Lagrangian is given by\footnote{We should note that the axion field $\chi$ is sourced by $F\wedge F$ and thus it is consistent to set it to zero only if $F\wedge F=0$. Since we are interested here in purely magnetic solutions, it is consistent to do so.}
\equ{
e^{-1}\cL=R \ast {\bf 1}-\frac12\ast d\phi\wedge d\phi-\frac12 e^{-\phi}\ast F^{I}\wedge F^{I}-\frac12 e^{\phi}\ast \tilde{F}^{I}\wedge \tilde{F}^{I}-V\ast {\bf 1}
}
where $I=1,2,3$ are $SU(2)$ indices and the scalar potential is given by  
\equ{
V=-4\bar g^{2}\,(2+\cosh \phi)\,.
}
We set $\bar g=1/\sqrt2$ in what follows to  normalize the scale of the AdS$_{4}$ vacuum to $L_{\text{AdS}_{4}}=1$.

We are interested in solutions with gauge fields excited only along the Cartan of the two $SU(2)$'s which leads to the following Ansatz:
\eqss{
ds^{2}&=e^{2f(r)}\,(-dt^{2}+dr^{2})+e^{2g(r)}\,ds^{2}_{\Sigma_{\fg}}\,,\\
F^{I}&=-\frac{\mathfrak n}{\sqrt 2}\, \delta^{I3} \, \dvol(\Sigma_{\fg})\,,\qquad \tilde F^{I}=-\frac{\widetilde{ \mathfrak n}}{\sqrt 2}\, \delta^{I3} \, \dvol(\Sigma_{\fg})\,.\
}
The BPS equations read:
\eqss{\label{4dN=4BPS}
e^{-f}f'&=-\frac12 \, (e^{-\frac{\phi}{2}}+e^{\frac{\phi}{2}})+\frac{e^{-2g}}{2} \,(\mathfrak n \, e^{\frac{\phi}{2}}+\widetilde{ \mathfrak n} \,e^{-\frac{\phi}{2}})\,,\\
e^{-f}g'&=-\frac12 \,(e^{-\frac{\phi}{2}}+e^{\frac{\phi}{2}})-\frac{e^{-2g}}{2} \,(\mathfrak n \, e^{\frac{\phi}{2}}+\widetilde{ \mathfrak n} \, e^{-\frac{\phi}{2}})\,,\\
e^{-f}\phi'&=- (e^{-\frac{\phi}{2}}-e^{\frac{\phi}{2}})+e^{-2g}\, (\mathfrak n \, e^{\frac{\phi}{2}}-\widetilde{ \mathfrak n} \, e^{-\frac{\phi}{2}})\,,\\
\mathfrak n+
\widetilde{ \mathfrak n}&=\kappa\,.
}
These BPS equations can also be obtained as a truncation of the well-studied $U(1)^{4}$ STU model of four-dimensional gauge supergravity by setting pairs of the four $U(1)$ gauge fields equal and two dilatons and all three axions to zero. This was analyzed in some detail in \cite{Benini:2015eyy}.\footnote{To be precise, these match the BPS  Equations (A.27) in  \cite{Benini:2015eyy}, setting $L_{1}=L_{2}\equiv e^{-\frac{\phi}{2}}$ and $L_{3}=L_{4}\equiv e^{\frac{\phi}{2}}$,  $\mathfrak{n}_{1}=\mathfrak n_{2}\equiv  \mathfrak n$, and $\mathfrak n_{3}=\mathfrak n_{4}\equiv \widetilde{ \mathfrak n}$.} 

To determine the parameter space for which regular black hole solutions exist it suffices to look at possible AdS$_{2}$ vacua, where the metric functions take the form \eqref{IRfp}. Then, the first two equations in \eqref{4dN=4BPS}  imply that $\phi=\phi_{0}$ is a constant and we obtain the set of algebraic equations
\eqss{
2\,e^{-f_{0}}&= (e^{-\frac{\phi_{0}}{2}}+e^{\frac{\phi_{0}}{2}})-e^{-2g_{0}}\, (\mathfrak n \, e^{\frac{\phi_{0}}{2}}+\widetilde{ \mathfrak n} \,e^{-\frac{\phi_{0}}{2}})\,,\\
0&=(e^{-\frac{\phi_{0}}{2}}+e^{\frac{\phi_{0}}{2}})+e^{-2g_{0}} \,(\mathfrak n \, e^{\frac{\phi_{0}}{2}}+\widetilde{ \mathfrak n} \, e^{-\frac{\phi_{0}}{2}})\,,\\
0&= (e^{-\frac{\phi_{0}}{2}}-e^{\frac{\phi_{0}}{2}})-e^{-2g_{0}} \,(\mathfrak n \, e^{\frac{\phi_{0}}{2}}-\widetilde{ \mathfrak n} \, e^{-\frac{\phi_{0}}{2}})\,.\\
}
Note that combining the last two equations it follows that 
\equ{
\mathfrak n=\widetilde{ \mathfrak n}=\frac{\kappa}{2}=-\,e^{2g_{0}}\,.
}
Thus, an AdS$_{2}$ vacuum is found only for a twist along the diagonal Cartan of $SU(2)_{C}\times SU(2)_{H}$. This coincides with the universal black hole solution of the 4d $\cN=2$ supergravity theory discussed in Section \ref{subsubsec:4dN2SG} but now derived as a solution of the 4d $\cN=4$ minimal supergravity. Indeed, one sees that the solution coincides with the solution \eqref{solN=2BH} of the minimally supersymmetric theory. The fact that there are no regular AdS$_{2}$ vacua preserving four supercharges is consistent with the fact that the topologically twisted index of the corresponding $\cN=4$ theory does not have an $N^{3/2}$ scaling, as discussed in Section~\ref{sec:Comments on 3d and 5d SCFTs}. 

\subsection{Uplifts to 10d and 11d}
\label{subsec:Uplifts}

As emphasized throughout the paper, a crucial aspect of our story is the fact that the gauged supergravity solutions presented above are universal. This manifests itself in two distinct ways. On the one hand, they are solutions of the minimal gauged supergravity in a given spacetime dimension with a given amount of supersymmetry, i.e., we have switched off any possible matter multiplets. This is the simple  holographic dual of the fact that in the SCFT the universal partial topological twists discussed in Section \ref{sec:Field theory} involve only the stress-energy tensor multiplet. On the other hand, these minimal supergravity solutions can be embedded into string and M-theory in infinitely many distinct ways by using various consistent truncation results in the literature. This is ultimately the core statement of universality of these constructions; a particular embedding in string or M-theory describes the twisted compactification of a particular dual SCFT, and the universal relations between QFT quantities derived above hold independently of the details of the SCFTs at hand. 

Let us briefly describe how these embeddings into ten and eleven-dimensional supergravity are realized. It was shown in \cite{Gauntlett:2006ai} and \cite{Gauntlett:2007ma} that any supersymmetric AdS$_5$ vacuum of 11d or type IIB supergravity admits a consistent truncation to minimal 5d $\cN=2$ gauged supergravity. The same also holds for any supersymmetric AdS$_4$ solution of 11d supergravity which can be consistently truncated to 4d minimal $\cN=2$ gauged supergravity as shown in \cite{Gauntlett:2007ma}.\footnote{There is also an embedding of the minimal 4d $\cN=2$ gauged supergravity in massive IIA supergravity as discussed in \cite{Guarino:2015jca,Guarino:2015vca}.} Since there are infinitely many such AdS$_4$ and AdS$_5$ solutions we arrive at the conclusion that the universal black string and black hole solutions presented in Section \ref{sec:N=2 supergravity} and Section \ref{subsubsec:4dN2SG} can be embedded in 10 and 11d supergravity, respectively, in infinitely many ways. We note that this does not apply only to the usual Freund-Rubin type solutions and one can also find realizations of the universal flows for warped AdS$_5$ and AdS$_4$ vacua of IIB and 11d supergravity as shown explicitly in \cite{Bobev:2014jva} and \cite{BMP} respectively. In \cite{Gauntlett:2007sm} and  \cite{Lu:1999bw}  similar consistent truncation results were also derived for the minimal 5d $\cN=4$ gauged supergravity discussed in Section \ref{sec:5d N=4 supergravity}. In \cite{Gauntlett:2007sm} it was shown that every supersymmetric AdS$_5$ vacuum of 11d supergravity which preserves 16 supercharges admits a consistent truncation to the minimal 5d $\cN=4$ gauged supergravity. The same holds true for the AdS$_5\times S^5$ solution (and its orbifolds preserving with 16 supercharges) in type IIB supergravity \cite{Lu:1999bw}.\footnote{We are not aware of any other AdS$_5$ vacua with 16 supercharges in type IIB supergravity. If such solutions exist it is reasonable to conjecture that they will admit a consistent truncation to the 5d $\cN=4$ minimal supergravity theory.} The fact that supersymmetric AdS$_7$ vacua with 16 supercharges of type IIA and 11d supergravity lead to a consistent truncation to the minimal 7d $\cN=2$ gauged supergravity of Section \ref{sec:Minimal 7d gauged supergravity} was shown in \cite{Passias:2015gya}. Finally, the solutions of Romans' minimal six-dimensional supergravity theory discussed in Section \ref{sec:6d  supergravity} can be embedded into massive IIA or type IIB supergraviy using the results of \cite{Cvetic:1999un} and \cite{Jeong:2013jfc}, respectively. All of these consistent truncation results for AdS vacua with 16 supercharges are nicely captured by the recent analysis in \cite{Malek:2017njj}. It was shown in \cite{Malek:2017njj}  that any AdS$_D$ solution of 10d or 11d supergravity admits a truncation to the respective minimal (i.e. containing only the gravity multiplet) gauged supergravity in $D$ dimensions.\footnote{We note that the results in \cite{Malek:2017njj} are somewhat implicit and do not immediately lead to convenient explicit uplift formulas from $D$ to 10 or 11 dimensions. } 

The upshot of this collection of supergravity results is that every $d$-dimensional SCFT (for $d\geq 3$) with a weakly coupled string/M theory dual captured by a supersymmetric AdS$_{d+1}$ vacuum of 10d or 11d supergravity admits a consistent description in terms of minimal gauged supergravity in $d+1$ dimensions. This in turn implies that every such SCFT also enjoys the universal RG flows across dimensions by a twisted compactification on $M_{d-p}$ and the holographic description of this RG flow is in terms of the supergravity domain wall solutions discussed in this section.

This observation is particularly useful in the case of flows between odd-dimensional SCFTs. A simple example of this is the uplift of the AdS$_{4}$ black hole solution of Section~\ref{sec:4d supergravity}  to eleven-dimensional supergravity and to massive IIA supergravity, recently discussed in  \cite{ABCMZ}. Using the universal relation \eqref{BHFS3univ} then leads to the microscopic counting of the entropy of a large class of AdS$_{4}$ black holes \cite{ABCMZ} both in M-theory and in massive IIA string theory. It would be interesting to apply this approach to the microscopic counting of the entropy of the various black brane solutions described in this work.  This requires a detailed understanding of the field theory quantity computing the corresponding entropy, which is currently lacking. One example that might be interesting to study is the AdS$_{5}$ black hole, with near-horizon geometry \eqref{AdS25dsugra} and entropy \eqref{S1H3BH}. In analogy to the AdS$_{4}$ case, a natural guess for the field theory quantity that should capture its entropy is the partition function of a 4d $\cN=2$ theory on $S^{1}\times \Bbb H^{3}$, with a partial topological twist on $\Bbb H^{3}$. We leave the exploration of this interesting question for future work.

\section{Discussion}
\label{sec:Discussion}

We have established universal relations between physical observables in SCFTs with a continuous R-symmetry, connected by RG flows across dimensions. The precise flows are triggered by a partial topological twist on a compact manifold along the exact UV superconformal R-symmetry. The underlying reason for this universality is the fact that the deformation in the UV amounts to coupling the omnipresent  stress-energy tensor multiplet of the SCFT  to background fields, namely a background metric and R-symmetry gauge field, and switching off any possible couplings to flavor symmetry currents.  

If the  SCFT$_{d}$ in the UV admits a weakly coupled AdS$_{d+1}$  gravity dual in string or M-theory, and if the compactification manifold admits a constant negative-curvature metric, we have provided ample evidence that the $p$-dimensional theory in the IR is also conformal and  admits a weakly coupled AdS$_{p+1}$ dual. The gravitational description provides an explicit realization of the RG flow across dimensions via a simple domain wall solution of gauged supergravity,  interpolating between the AdS$_{d+1}$ vacuum in the UV and an AdS$_{p+1}$  vacuum in the IR. The universality of such flows is understood holographically by the fact that these domain walls can be uplifted to string or M-theory in infinitely many ways.

Our results suggest various interesting questions and directions for future work.  Clearly, the most pressing and general question is how to find an independent description of the low-energy $p$-dimensional SCFTs. We have defined these theories via a twisted compactification of higher-dimensional theories. It would be  valuable, however, to have a UV definition in terms of a theory living in the same number of spacetime dimensions. This can be achieved either through a direct definition of the CFT, e.g., by a nonlinear $\sigma$-model on a Ricci-flat target manifold in the case $p=2$, or via some $p$-dimensional asymptotically free UV description that flows to the interacting CFT in the IR. The gold standard for this is set by the $\mathcal{N}=1$ and $\mathcal{N}=2$ theories of class $\mathcal{S}$ and the 3d theories of class $\mathcal{R}$ arising from M5-branes wrapping Riemann surfaces and three-manifolds, respectively--see \cite{Gaiotto:2009we,Dimofte:2011ju} and references thereof. Perhaps the most accessible setup to generalize this success to other dimensions is to focus on $d=4,\,p=2$, in particular the $\alpha$- and $\beta$-twists of four-dimensional $\mathcal{N}=2$ theories  discussed in Section~\ref{sec:N=2 on Sigma} and holographically in Section~\ref{sec:5d N=4 supergravity}. Some progress in this direction was made recently in \cite{Putrov:2015jpa,Gukov:2017zao}, but there is certainly more to be understood, especially for theories in the large $N$ limit.

When the RG flow  is between even-dimensional SCFTs we relied purely on field theory methods, in particular the power of 't Hooft anomaly matching and superconformal symmetry, to derive exact,  finite $N$, relations between quantities in the UV and IR theories. As shown in Section \ref{sec:Gauged Supergravity} these relations are reproduced holographically to leading order in $N$, thus providing strong evidence for the existence of such flows and IR fixed points. The supergravity analysis, however, is not limited to flows between even-dimensional SCFTs and we have used properties of various supergravity solutions to predict similar universal relations when one (or both) of the SCFTs is odd-dimensional, in which case the appropriate physical quantity is the round-sphere free energy. These relations, as currently stated in Section \ref{sec:Gauged Supergravity}, are established only in the large $N$ limit. It would be most interesting to study whether this picture extends beyond the planar limit. Ideally, this could be approached by an exact field theory calculation using supersymmetric localization on the appropriate curved manifold as, e.g., the case of three-dimensional theories on Riemann surfaces \cite{Benini:2015eyy,Hosseini:2016tor,ABCMZ}. It is likely that this can be generalized further by studying, for instance, suitable supersymmetric partition functions of five-dimensional SCFTs with a partial topological twist on $ \Sigma_{\fg}$ to reproduce the holographic prediction in \eqref{FS3FS5} by pure field theory methods. Similarly, it should be possible to study four-dimensional $\mathcal{N}=2$ SCFTs with a topological twist on  $M_3$, with $M_3$ an appropriate hyperbolic manifold, to reproduce the black hole entropy in \eqref{S1H3BH}.  An alternative approach to incorporating subleading corrections in $N$ would be to analyze the universal RG flows holographically,  including higher-curvature corrections in gauged supergravity. Although a technically challenging problem in general, this was addressed successfully in \cite{Baggio:2014hua} for various domain walls interpolating between AdS vacua corresponding to RG flows between even-dimensional SCFTs. It would be very interesting to extend this approach to the various domain wall solutions described here.  It would also be interesting to study whether there are universal relations among other physical observables in the $p$-dimensional and $d$-dimensional SCFTs, e.g., Wilson loop expectation values or partition functions with other insertions.  

A series of interesting questions relate to the choice of $M_{d-p}$. As discussed, most universal RG flows we have studied, both in field theory as well as holographically, require $M_{d-p}$ to be hyperbolic or negatively curved in order for the IR $p$-dimensional theory to be unitary.\footnote{An exception to this general rule is the case of 6d SCFTs on a 5-manifold at large $N$, as noted in Section~\ref{sec:maximal 7d supergravity}. We have assumed throughout the paper that the UV $d$-dimensional theory is unitary.} We do not have an explanation why this must be the case in general and it would be interesting to have a better understanding of this. In addition, the supergravity solutions we constructed require the metric on the compact manifold $M_{d-p}$ in the IR to be Einstein. From field theory considerations, however, it is clear that in the UV one should be able to use any metric on $M_{d-p}$, since we are performing a topological twist. Thus, holography  suggests that the RG flow across dimensions uniformizes the metric on $M_{d-p}$. This has been understood in some detail for Riemann surfaces in \cite{Anderson:2011cz} and it would certainly be very interesting to explore the interplay between holographic RG flows and uniformization for higher-dimensional manifolds. The Einstein metric at the IR end of the RG flow may still admit a moduli space of deformations compatible with the Einstein condition. These moduli should correspond to exactly marginal couplings in the $p$-dimensional SCFT. This is also clear from the gravitational construction of the holographic dual, where the moduli of the Einstein metric on $M_{d-p}$ lead to massless scalar excitations on the AdS$_{p+1}$ space dual to the SCFT. This picture is well-established for the case $d=6,\, p=4$ \cite{Gaiotto:2009we,Benini:2009mz,Bah:2011vv,Bah:2012dg} and partially explored for the case $d=6,\, p=2$ in \cite{Gadde:2013sca}. Additional exactly marginal deformations of the $p$-dimensional SCFT may be present if the $d$-dimensional parent theory admits global, non-R, symmetries for which one can turn on flat connections on $M_{d-p}$. Finally, let us note that we have assumed throughout the paper that  $M_{d-p}$  is smooth and compact. It is natural to consider generalizations of this setup to allow for boundaries, punctures, or other defects. It would be interesting to study whether there is a generalization of the universal relations uncovered in this work in these more general situations.

As discussed at length above, the main reason behind the existence of the universal RG flows across dimensions is that the partial topological twist  triggering the RG flow is performed using only background fields that couple to the universal stress-energy tensor multiplet, which exists for all SCFTs with a continuous R-symmetry. The holographic manifestation of this universality is realized by the fact that gauged supergravities always admit a truncation to a universal sector  including only the gravity multiplet. In addition, these $(d+1)$-dimensional ``minimal'' supergravities arise as universal consistent truncations from string and M-theory in infinitely many ways, that are distinguished by the choice of internal manifold (and the fluxes through it) used for the reduction from 10 (or 11) dimensions. This holographic perspective suggests that SCFTs with a holographic dual enjoy a truncation of the OPE, at least at large $N$, for operators belonging to the stress-energy multiplet. It would be very interesting to understand the mechanism behind such an OPE truncation, as this could offer an explanation and organizational principle for the plethora of consistent truncations in the supergravity literature. 

A few other observations made throughout this work deserve further analysis. What is the relation between unitarity of the IR $p$-dimensional theory at the end of the RG flow across dimensions and the Hofman-Maldacena-type bounds for the parent $d$-dimensional UV theory? Is there a Hofman-Maldacena-type bound on the four anomaly coefficients in six-dimensional SCFTs?\footnote{See for example \cite{Zhou:2015kaj,Zhou:2016kcz,Yankielowicz:2017xkf} for some recent results and conjectures in this direction.} 

From all the examples of RG flows across dimensions studied here, it seems that to obtain an SCFT with an AdS dual in the IR, the UV theory must also be conformal and strongly interacting. We are not aware of any a priori reason for this to be the case, and it would be interesting to find more general examples of RG flows across dimensions, where the UV theory is not strongly interacting. In Appendix~\ref{app:entropy} we observe intriguing relations between field theory observables such as conformal anomalies and free energies and the entropies of various supersymmetric black branes. It is desirable to put these on a firmer footing and calculate the black brane entropies from a more rigorous field theory setting, as done recently for supersymmetric black holes in AdS$_4$ \cite{Benini:2015eyy,ABCMZ}. Finally, it is natural to wonder whether there is some notion of a ``monotonicity theorem,'' similar to the $c$-, $a$-, or $F$-theorems for RG flows across dimensions. Based on the universal relations between conformal anomalies and free energies studied in this work it is clear that simply comparing the natural monotonic quantity in the IR $p$-dimensional SCFT with the one in the UV $d$-dimensional SCFT is too naive. Perhaps one should search for a more refined definition of a monotonic function along the RG flow which removes the explicit factor of the volume of the compactification manifold $M_{d-p}$.

It is clear that the universal RG flows described here provide a fertile area for exploring the physics of supersymmetric QFTs and holography. We expect many further exciting developments ahead of us.

\bigskip
\bigskip

\textbf{Acknowledgments}

\medskip

\noindent We are grateful to Fabio Apruzzi, Marco Baggio, Chris Beem, Davide Cassani, Eoin \'O Colg\'ain, Ant\'on Faedo, Fri\dh rik Gautason, Diego Hofman, Edoardo Lauria, Vincent Min, Krzysztof Pilch, Alberto Zaffaroni, and especially Francesco Benini for interesting discussions. The work of NB is supported in part by the starting grant BOF/STG/14/032 from KU Leuven, by an Odysseus grant G0F9516N from the FWO, and by the KU Leuven C1 grant ZKD1118 C16/16/005. In addition NB acknowledges support by the Belgian Federal Science Policy Office through the Inter-University Attraction Pole P7/37, and by the COST Action MP1210 The String Theory Universe. PMC is supported by Nederlandse Organisatie voor Wetenschappelijk Onderzoek (NWO) via a Vidi grant. The work of PMC is part of the Delta ITP consortium, a program of the NWO that is funded by the Dutch Ministry of Education, Culture and Science (OCW). PMC would like to thank the ITP at Stanford University and KU Leuven for kind hospitality while part of this work was carried out. Both of us are grateful for the warm hospitality offered by the CERN Theory Group at the final stages of this work.

\appendix

\section{Conventions and normalizations}

In this appendix we review our conventions and normalizations and collect useful formulae used throughout the paper.

\subsection{Characteristic classes}
\label{App:CC}

The total Chern class of a vector bundle and the Pontryagin class of the tangent bundle are given by  (here we are following the conventions in  \cite{Harvey:2005it}, see also \cite{AlvarezGaume:1983ig})
\eqsn{
C(\cF)&=\det\(1+i\frac{\cF}{2\pi}\)=1+c_{1}(\cF)+c_{2}(\cF)+\cdots\,, \\
P(\cT)&=\det\(1-\frac{\cR}{2\pi}\)=1+p_{1}(\cT)+p_{2}(\cT)+\cdots\,,
}
with $\cF$ the field strength two-form, which we take to be antihermitian, and $\cR$ the Riemann curvature two-form. From this we find the following expressions for the first two Chern and Pontryagin classes
\eqs{
c_{1}(\cF)&=\frac{i}{2\pi}\, \Tr \cF\,, &&  c_{2}(\cF)=\frac{1}{(2\pi)^{2}}\,\frac12\,\(\Tr \cF^{2}-(\Tr \cF)^{2}\)\,,\\
p_{1}(\cT)&=-\frac{1}{(2\pi)^{2}}\,\frac12\, \Tr \cR^{2}\,, && p_{2}(\cT)=\frac{1}{(2\pi)^{4}}\,\( \frac18 \,(\Tr \cR^{2})^{2}-\frac14\,\Tr \cR^{4}\)\,,
}
where in the first line the $\Tr$ is over ``gauge'' indices in the fundamental representation and in the second line the $\Tr$ is over tangent frame indices. Note that $p_{1}(\cT)$ differs by a sign from that used in \cite{Cordova:2015fha} which explains the minus sign difference in $\beta$ in \eqref{I8} in comparison to the expression in \cite{Cordova:2015fha}. 

A useful property of characteristic classes is the splitting principle. 
The total Chern and Pontryagin classes  $C(U)=\sum_{i}c_{i}(U)$, $P(E)=\sum_{i}p_{i}(E)$, decompose under the direct sum of vector bundles as
\equ{
C(U\oplus V)=C(U) \,C(V)\,,\qquad P(E\oplus F)=P(E) \,P(F)\,.
}
In particular, it follows from the second relation that the first and second Pontryagin classes satisfy
\equ{\label{p1p2M2M4}
p_{1}(E\oplus F)=p_{1}(E)+p_{1}(F)\,,\qquad p_{2}(E\oplus F)=p_{2}(E)+p_{2}(F)+p_{1}(E)\,p_{1}(F)\,,
}
and similarly for  Chern classes.
Given a decomposition of the tangent bundle as a sum of complex line
bundles with first Chern classes $e_{i}$ one has:
\equ{
p_{1}(\cT)=\sum_{i} e_{i}^{2}\,,\qquad p_{2}(\cT)=\sum_{i<j} e_{i}^{2}\,e_{j}^{2}\,.
}
%
%
Finally we note that for a four-manifold which is a product of two Riemann surfaces $\Sigma_{1}\times \Sigma_{2}$, the integrated first Pontryagin class, $P_{1}$, and the Euler characteristic, $\chi$,  are $P_1=0$ and $\chi = 4(\fg_1-1)(\fg_2-1)$.

\subsection{Metric on Riemann surfaces}

Throughout this paper we often consider smooth Riemann surfaces $\Sigma_{\fg}$ of genus $\fg$. We always put a constant curvature metric on these manifolds with the following explicit form 
\begin{equation}\label{hdef}
ds^{2}_{\Sigma_{\fg}}=e^{2h(x_1,x_2)}\,(dx_{1}^{2}+dx_{2}^{2})\,,\quad h(x_1,x_2) = \begin{cases} -\log \frac{1 + x_1^2 +x_2^2}2 & \text{for } \fg = 0 \\ \frac12 \log 2\pi & \text{for } \fg = 1 \\ - \log x_2 & \text{for } \fg >1\;\end{cases} \,.
\end{equation}
The volume form $\dvol_{\Sigma_{\fg}}\equiv e^{2h}\,dx_{1}\wedge dx_{2}$ integrates to:
\eqs{\label{norm tg}
\int \dvol(\Sigma_{\fg})=2\pi\eta_{\Sigma} \,, \qquad \eta_{\Sigma} =\begin{cases}
               2|\fg-1| &\qquad\text{for $\fg\neq 1$}\\
              1 &\qquad\text{for $\fg= 1$}
            \end{cases}\,.
}
The normalized  curvature of $\Sigma_{\fg}$ is denoted  by $\kappa=\{1,0,-1\}$ for $\fg=0$, $\fg=1$, and $\fg>1$, respectively.  We note that with these definitions and using \eqref{norm tg} one has the relation $\kappa \, \eta_{\Sigma}=-2(\fg-1)$ for all $\fg$.  Finally,  $t_{\fg}$ denotes the first Chern class of the tangent bundle of $\Sigma_{\fg}$, which in our normalizations integrates to $\int_{\Sigma_{\fg}}t_{\fg}=\eta_{\Sigma}$.

\section{Entropy of black branes}
\label{app:entropy}

All regular supergravity solutions discussed in this paper can be viewed as extremal $(p-1)$-brane solutions in $(d+1)$ spacetime dimensions. The near horizon geometry is of the form:
\equ{\label{sol IR app}
ds^{2}_{d+1}=e^{2f_{0}}\, ds^{2}_{\text{AdS}_{p+1}}+e^{2g_{0}}\, ds^{2}_{M_{d-p}}\,,
}
where 
\begin{equation}
ds^{2}_{\text{AdS}_{p+1}}=\frac{1}{r^{2}}(-dt^{2}+dr^{2}+dz_{1}^{2}+\cdots+dz_{p-1}^{2})\,,
\end{equation}
and $ds^{2}_{M_{d-p}}$ is the metric on the compact horizon of the $(p-1)$-brane. The field theory interpretation of these $(p-1)$-brane solutions is given by the universal RG flows across dimensions discussed extensively in the main text. In particular, there is a $p$-dimensional SCFT captured holographically by the AdS$_{p+1}$ factor in the near horizon geometry. This is the IR SCFT which arises from a $d$-dimensional UV SCFT via the RG flow across dimensions. For $p$ even, the conformal anomaly coefficients of the SCFT can be computed holographically using  \eqref{achol}.  Another physically interesting quantity is the entropy density of the black brane. To compute this we take the spatial coordinates on the boundary of $\text{AdS}_{p+1}$, $z_{i}$, $i=1,\cdots,p-1$, to have a finite range $z_{i}\in [0,l_{i}]$.  We can then easily compute the Bekenstein-Hawking entropy per unit of spatial volume, $V\equiv l_{1} \times \ldots \times l_{p-1}$,\footnote{We note that this is the spatial volume of the boundary of $\text{AdS}_{p+1}$ and not that of the horizon $M_{d-p}$.}
\equ{
s\equiv\frac{S}{V}=\frac{\vol(\tilde M_{d-p})\,e^{(p-1)f_{0}}}{4G_{N}^{(d+1)}}=\frac{e^{(p-1)f_{0}}}{4G_{N}^{(p+1)}}\,,
}
where $ds^{2}_{\tilde M_{d-p}}\equiv e^{2g_{0}}ds^{2}_{M_{d-p}}$ and in the last equality we used the relation between the Newton constants in $p+1$ and $d+1$ dimensions, 
\begin{equation}
\ds\frac{1}{G_{N}^{(d+1)}}=\ds\frac{\vol(\tilde M_{d-p})}{G_{N}^{(p+1)}}\,.
\end{equation}
Combining this with the formulae   \eqref{achol}  for the central charges in a $p$-dimensional SCFT for $p=2,4,6$ we can write the entropy density of a $(p-1)$-brane in terms of the corresponding Weyl anomaly coefficient:
\eqs{\label{entp D1}
s(D1)&=\frac{1}{6}\,c_{2d}\,,\\ \label{entp D3}
s(D3)&=\frac{2}{\pi}\,a_{4d}\,,\\  \label{entp D5}
s(D5)&=\frac{7}{12\pi^{2}}\,a_{6d} \,.
}
In the first line we defined $c_{2d}\equiv c_{r}=c_{l}$, which is valid in the large $N$ limit. 

Similarly, for $p=3,5$ we can write the entropy density in terms of the corresponding round-sphere free energy  in \eqref{Fgrav S3} for the SCFT$_{p}$, in the following way
\eqs{\label{entp D2}
s(D2)&=\frac{F_{S^{3}}}{2\pi}\,, \\  \label{entp D4}
s(D4)&=-\frac{3F_{S^{5}}}{4\pi^{2}}\,.
}
We emphasize that the anomaly coefficients and free energies appearing in \eqref{entp D1}-\eqref{entp D4} are those corresponding to the SCFT living on the worldvolume of the $(p-1)$-brane, i.e., the IR theory. The universal relations between conformal anomalies and free energies in the large $N$ limit that we discussed in this work  allow us to relate these IR observables to UV quantities corresponding to the SCFT$_{d}$ dual to the asymptotically locally $\text{AdS}_{d+1}$ boundary of the black brane solution. 

Finally, we point out that the case $p=1$, i.e., an AdS$_{2}$ near horizon region, is somewhat special since we do not have a microscopically well-established $\text{AdS}_{2}/\text{CFT}_1$ duality. The best understood setup is that of extremal black holes in $\text{AdS}_{4}$ discussed in Section \ref{sec:4d supergravity}.

\section{Universal RG flows in the same dimension}
\label{app:universal2d2d}

While the main focus in this work has been to uncover universal relations between SCFTs connected by an RG flow across dimensions, it is important to note that there are also similar relations for SCFTs living in the same number of spacetime dimensions. A particularly simple relation between the conformal anomalies of 4d $\cN=2$ and $\cN=1$ SCFTs connected by a specific universal RG flow was derived by Tachikawa and Wecht in \cite{Tachikawa:2009tt}. Inspired by this here we seek a similar result  for 2d SCFTs.

First, let us revisit the main result in \cite{Tachikawa:2009tt} from the perspective of our discussion. A particular class of examples of 4d $\cN=2$ and  $\cN=1$ SCFTs which are connected by the RG flow of \cite{Tachikawa:2009tt} are the MN $\cN=2$ and MN $\cN=1$ SCFTs discussed in Section \ref{subsec:6dN20CFT} above. The relation between the anomalies of these two classes of 4d SCFTs can be obtained by using the two universal relations, \eqref{ac4dn=120} and \eqref{matMN2}, of these anomalies to the ones of the class of $\cN=(2,0)$ SCFTs in 6d. Using \eqref{ac4dn=120} and inverting the matrix in \eqref{matMN2} we find that the 4d $\cN=2$ and $\cN=1$ central charges are related by
\eqs{\label{TW4d}
\begin{pmatrix}
a_{4d}\\
c_{4d}
\end{pmatrix}_{\cN=1}\,=\,\,\,\,\,\, \frac{3}{32}\begin{pmatrix}
12& -3\\
-4&13
\end{pmatrix} \begin{pmatrix}
a_{4d}\\
c_{4d}
\end{pmatrix}_{\cN=2}\,.
}
This is precisely the relation derived in \cite{Tachikawa:2009tt}.

We can now apply the same idea for the $\alpha$- and $\beta$-twists discussed in Section \ref{sec:N=2 on Sigma} and the universal $(0,2)$ twist of Section \ref{N=1 Sigma}. For the $\alpha$-twist the matrix relating the 2d and 4d anomaly coefficients is not invertible and thus we cannot derive the analog of a Tachikawa-Wecht result. For the $\beta$-twist, however, the matrix in \eqref{clcr2004B} is invertible and one can combine that with the matrix in \eqref{universalmat}  to find the following 2d analog of the Tachikawa-Wecht relation
\eqs{\label{TW2d}
\begin{pmatrix}
c_{r}\\
c_{l}
\end{pmatrix}_{(0,2)}\,=\,\,\,\,\,\frac{1}{9}\begin{pmatrix}
5& -1\\
2&2
\end{pmatrix} \begin{pmatrix}
c_{r}\\
c_{l}
\end{pmatrix}_{(0,4)}\,.
}
We also observe that, as discussed in Section \ref{sec:N=34 on Sigma}, for 4d $\cN=3$ theories one can have universal flows to 2d preserving $\cN=(0,6)$ and $\cN=(2,4)$ supersymmetry. Comparing the central charges \eqref{N=3to2,4} and \eqref{N=3to0,6} to those obtained for the universal $\cN=(0,2)$ twist \eqref{universalmat}, and using that  $a_{4d}=c_{4d}$ for 4d $\cN=3$ SCFTs, one obtains the relations
\begin{equation}
\label{TW4dN3}
c_{\,(0,2)} = \frac{5}{9}\, c_{\,(0,6)}\;, \qquad c_{\,(0,2)} = \frac{4}{9} \,c_{\,(2,4)}\;.
\end{equation}
Here we have denoted $c_{r}=c_{l}\equiv c$, which holds for these 2d theories.  The second equation above is a special case of the relation in \eqref{TW2d}, applied to $\cN=(0,4)$ theories with $c_r=c_l$. We also note that the $\beta$-twist, applied to $\cN=4$ SYM leads to a 2d theory with enhanced $\cN=(4,4)$ and central charge $c_{\,(4,4)} = 6(\fg-1)d_G$, with $d_{G}$ the dimension of the gauge group,  while the universal $(0,2)$ twist applied to the $\cN=4$ theory leads to a 2d SCFT with central charge $c_{\,(0,2)} = \frac{8}{3} (\fg-1)d_G$, consistent with the second equation in \eqref{TW4dN3}. See \cite{Benini:2013cda} for a detailed discussion of partial topological twists of $\cN=4$ SYM. 

We emphasize that although the relations in \eqref{TW2d} and \eqref{TW4dN3} are very suggestive, we have not derived them by studying concrete two-dimensional RG flows realized in a specific SCFT. It would be most interesting to explore this further and understand whether such universal two-dimensional RG flows indeed exist.


\bibliography{References} 
\bibliographystyle{JHEP}

\end{document}